\documentclass[twocolumn]{aastex63}
\usepackage{latexsym}
\usepackage{amssymb}
\usepackage{amsfonts}
\usepackage{amsmath}
\usepackage{bm}
\usepackage{tabularx}
\usepackage[normalem]{ulem}
\usepackage{multirow}
\usepackage{natbib} 
\usepackage{gensymb}
\usepackage{subfigure}
\usepackage{wrapfig}
\usepackage{color}
\usepackage{calc}
\usepackage{amsmath,amssymb,graphicx}
\usepackage{amssymb,amsmath}
\usepackage{tensor}
\usepackage{bm}
\usepackage{pifont}
\usepackage{microtype}
\usepackage{booktabs}
\usepackage{times}
\usepackage[varg]{txfonts}
\usepackage{subfigure}
\usepackage{lipsum}
\usepackage{breakurl} 
\usepackage{graphicx}
\usepackage{pbox}

\definecolor{LinkColor}{rgb}{0.75, 0, 0}
\definecolor{CiteColor}{rgb}{0, 0.5, 0.5}
\definecolor{UrlColor}{rgb}{0, 0, 0.75}
\hypersetup{linkcolor=LinkColor}
\hypersetup{citecolor=CiteColor}
\hypersetup{urlcolor=UrlColor}
\allowdisplaybreaks
\DeclareFontFamily{OT1}{pzc}{}
\DeclareFontShape{OT1}{pzc}{m}{it}{<-> s * [1.10] pzcmi7t}{}
\DeclareMathAlphabet{\mathpzc}{OT1}{pzc}{m}{it}

\newcommand{\caps}[1]{{\scshape{#1}}}
\newcommand{\hst}{\textit{HST}}
\newcommand{\xmm}{\textit{XMM-Newton}}
\newcommand{\nustar}{\textit{NuSTAR}}
\def\psr{PSR~J1023$+$0038}
\def\fgl{3FGL~J1544.6$-$1125}
\def\xss{XSS~J1227$-$4853}
\def\igr{IGR~J1824$-$2452}
\def\kep{\textit{Kepler}}

\def\kms{km s$^{-1}$}

\def\kms{km s$^{-1}$}

\submitjournal{ApJ}

\shorttitle{Multiwavelength study of \psr}
\shortauthors{Jaodand et al.}

\begin{document}

\title{Discovery of UV millisecond pulsations and moding in the low mass X-ray binary state of transitional millisecond pulsar J1023+0038}

\correspondingauthor{Amruta Jaodand}
\email{ajaodand@caltech.edu}

\author[0000-0002-3850-6651]{Amruta D. Jaodand}
\affil{ASTRON, Netherlands Institute for Radio Astronomy, Oude
Hoogeveensedijk 4, 7991 PD Dwingeloo, The Netherlands}
\affiliation{Anton Pannekoek Institute for Astronomy, University of Amsterdam, Science Park 904, 1098 XH, Amsterdam, The Netherlands}
\affiliation{California Institute of Technology, 1200 E Califronia Blvd., Pasadena, CA 91125, USA}

\author[0000-0002-6733-5556]{Juan V. Hern\'andez Santisteban}
\affil{Anton Pannekoek Institute for Astronomy, University of Amsterdam, Science Park 904, 1098 XH, Amsterdam, The Netherlands}
\affiliation{SUPA School of Physics \& Astronomy, University of St Andrews, North Haugh, St Andrews KY16 9SS, UK}

\author[0000-0003-0638-3340]{Anne M. Archibald}
\affiliation{Anton Pannekoek Institute for Astronomy, University of Amsterdam, Science Park 904, 1098 XH, Amsterdam, The Netherlands}

\author[0000-0003-2317-1446]{Jason W.~T. Hessels}
\affil{ASTRON, Netherlands Institute for Radio Astronomy, Oude
Hoogeveensedijk 4, 7991 PD Dwingeloo, The Netherlands}
\affiliation{Anton Pannekoek Institute for Astronomy, University of Amsterdam, Science Park 904, 1098 XH, Amsterdam, The Netherlands}

\author[0000-0002-9870-2742]{Slavko Bogdanov}
\affiliation{Columbia Astrophysics Laboratory, Columbia University, 550 West 120th Street, New York, NY 10027, USA
}

\author[0000-0002-1116-2553]{Christian Knigge}
\affiliation{School of Physics \& Astronomy, University of Southampton, Southampton SO17 1BJ, UK
}

\author[0000-0002-0092-3548]{Nathalie Degenaar}
\affiliation{Anton Pannekoek Institute for Astronomy, University of Amsterdam, Science Park 904, 1098 XH, Amsterdam, The Netherlands}

\author[0000-0001-9434-3837]{Adam T. Deller}
\affiliation{Centre for Astrophysics and Supercomputing, Swinburne University of Technology, John St, Hawthorn, VIC 3122, Australia}

\author[0000-0001-5387-7189]{Simone Scaringi}
\affiliation{Department of Physics and Astronomy, Texas Tech University, Lubbock, TX 79409-1051, USA}

\author[0000-0002-6459-0674]{Alessandro Patruno}
\affiliation{Leiden Observatory, Leiden University, P.O. Box 9513, NL-2300 RA Leiden, The Netherlands}
\affiliation{Institute of Space Sciences (IEEC-CSIC) Campus UAB, Carrer de Can Magrans, s/n, E-08193 Barcelona, Spain}



\begin{abstract}

\psr\ is a rapidly-spinning neutron star with a low-mass-binary companion that switches between a radio pulsar and low-luminosity disk state. In 2013, it transitioned to its current disk state accompanied by brightening at all observed wavelengths. In this state, \psr\ now shows optical and X-ray pulsations and abrupt X-ray luminosity switches between discrete `low' and `high' modes. Continuum radio emission, denoting an outflow, is present and brightens during the X-ray low modes. We present simultaneous optical, ultraviolet (UV) and X-ray campaign comprising \kep\ ($400-800$\,nm), \textit{Hubble Space Telescope} ($180-280$\,nm), \textit{XMM-Newton} ($0.3-10$\,keV) and \nustar\ ($3 - 79$\,keV). We demonstrate that low and high luminosity modes in the UV band are strictly simultaneous with the X-ray modes and change the UV brightness by a factor of $\sim25$\% on top of a much brighter persistent UV component. We find strong evidence for UV pulsations (pulse fraction of $0.82\pm0.19$\%) in the high-mode, with a similar waveform as the X-ray pulsations making it the first known UV millisecond pulsar. Lastly, we find that the optical mode changes occur synchronously with the UV/X-ray mode changes, but optical modes are inverted compared to the higher frequencies. There appear to be two broad-band emission components: one from radio to near-infrared/optical that is brighter when the second component from optical to hard X-rays is dimmer (and vice-versa). We suggest that these components trace switches between accretion into the neutron star magnetosphere (high-energy high-mode) versus ejection of material (low-energy high-mode). Lastly, we propose that optical/UV/X-ray pulsations can arise from a shocked accretion flow channeled by the neutron star's magnetic field.

\end{abstract}

\keywords{Ultraviolet sources--Accretion--Pulsars--Millisecond pulsars--Binary pulsars--Low-mass x-ray binary--Optical pulsars}



\section{Introduction} \label{sec:intro}


Transitional millisecond pulsars (tMSPs) are a class of neutron star binary that has emerged in the last decade with the discoveries of three systems: \psr~\citep{ABP:2015}, \xss~\citep{deMartino:2010, BPH:2014} and \igr~in the globular cluster M28 \citep[also known as M28I;][]{Papitto:2013}. These systems switch between a radio millisecond pulsar (RMSP) state and another state, where the system is brighter at most observed wavelengths. In this second state, these systems show an accretion disk, and give the appearance of low-level accretion into the neutron star magnetosphere. We therefore refer to this state as the low-mass X-ray binary (LMXB) state in this paper. However, we caution the reader that the ``LMXB state'' we refer to here is atypical of LMXB systems in general\footnote{Hence, we also refer to the LMXB state as the ``disk state'' although the nature and even existence of accretion onto the star in this state remains in doubt.  Other papers in the literature have referred to this state as the ``sub-luminous accretion disk state''. }, and \igr\ is the only tMSP system observed to enter a high X-ray luminosity ($> 10^{36}$\,erg\,s$^{-1}$) state that resembles a canonical LMXB in outburst.  

Sudden transitions between RMSP and LMXB states occur on a time scale of a few days to weeks, and are accompanied by drastic changes across the electromagnetic spectrum. For example, the transition from RMSP to LMXB state is accompanied by brightening of optical, UV \citep{Papitto:2013, TLK:2014, PAH:2014}, X-ray and $\gamma$-ray \citep{SAB:2013} emissions along with the simultaneous disappearance of radio pulsations\citep{PAH:2014}.  What drives these state transitions and the observed phenomena is still debated.  Intense multi-wavelength monitoring campaigns have been undertaken in the last decade to understand the rich observational phenomenology in both the RMSP and LMXB state.  \psr\ and the candidate tMSP \fgl\ \citep{BH:2015} are currently in the LMXB state, whereas \xss\ and \igr\ are currently in the RMSP state. In the following subsections, we provide a brief overview of observational characteristics that define the tMSP class, but caution that given the paucity of sources, it remains to be demonstrated how reproducible the multi-wavelength phenomenology is from source to source. \psr\ has been particularly informative because of its proximity \citep[$d = 1.37$\,kpc;][]{DAB:2012} and prolonged, ongoing LMXB state since.


\subsection{X-ray emission}
While in the RMSP state, the tMSPs show an orbitally modulated X-ray emission with $1-10$\,keV luminosity $L_{\rm X}<10^{32}$\,erg s$^{-1}$ \citep[for \psr, see][]{AKB:2010,BAH:2011}. This emission is thought to originate from an intra-binary shock \citep{BGV:2005} caused by the pulsar wind ramming into the companion's stellar wind near the surface of the companion. Similar behaviour has been observed \citep{HuiB:2006} in `redback' MSPs, which are short-orbital-period ($4-15$\,hr), RMSP binaries with a low-mass (${\sim}0.15$--$0.4 M_{\odot}$), non-degenerate companion \citep{Mal:2013}. In fact, all observed tMSPs so far belong to the redback class of RMSPs when they are in the radio pulsar state. 

Upon transition from the RMSP to the LMXB state, the orbitally modulated X-ray emission disappears and is replaced by a much brighter (up to $L_{\rm X} \sim 10^{33}$\,erg s$^{-1}$) X-ray emission. This X-ray luminosity is still low compared to an LMXB in outburst, however, and would be considered `quiescent' for a canonical LMXB source \citep{WDP:2017}.  Though on average the X-ray luminosity is stable over multi-year timescales \citep{ABP:2015, TYK:2014, JAH:2016}, the X-ray lightcurve is characterised by random, abrupt (within $\lesssim 10$\,s) switches between a steady `high' and `low' mode.  For \psr, the high mode is at $L_{\rm X} \sim 10^{33}$\,erg s$^{-1}$ and is present ${\sim} 80\%$ of the time, and the low mode is at $L_{\rm X} \sim 10^{32}$\,erg s$^{-1}$ and is present for $\sim 20\%$ of the time. There are also sporadic flares reaching up to $L_{\rm X} \sim 10^{34}$\,erg s$^{-1}$. Coherent X-ray pulsations at the pulsar's spin period are observed, but only in the `high' mode \citep[strong upper limits were obtained in the other modes;][]{ABP:2015}.

\subsection{$\gamma$-ray emission}
In the RMSP state, the $\gamma$-ray emission from tMSPs has been seen to be pulsed at the spin period of the pulsar \citep{AKH:2013, JRR:2015} -- similar to other \textit{Fermi}-Large Area Telescope (\textit{Fermi}-LAT) detected RMSPs.  For \psr, the $\gamma$-ray brightness is enhanced by a factor of ${\sim} 5$ \citep{SAB:2013} in the LMXB state.  \textit{Fermi}-LAT $\gamma$-ray ($\sim$GeV energies) pulsation searches in the LMXB state are limited by stochastic orbital variations that are not modeled in the rotational ephemeris \citep{JAH:2016}; this complicates the detection of $\gamma$-ray pulsations and currently no such pulsations have been seen in the LMXB state using \textit{Fermi}-LAT (Jaodand et al., \textit{in prep.}). Likewise, $\gamma$-ray pulsation searches using the Very Energetic Radiation Imaging Telescope Array System \citep[\textit{VERITAS}, spanning 50\,GeV$-$50\,TeV;][]{Ver:2016} have also found no significant $\gamma$-ray pulsations above $100$\,GeV. 

\subsection{UV and Optical emission}
Optical emission in the RMSP state is seen to be orbitally modulated: the optical emission peaks when the companion is at superior conjunction with respect to the pulsar.  This is due to the intra-binary shock and companion surface being heated by deposition of pulsar wind \citep{deMartino:2013,McConnell:2015}. Absorption lines from the companion are also observed and are used to identify the spectral type. 

In the LMXB state, the optical emission increases and a double peaked, broad H$\alpha$ emission line shows that an accretion disk with orbital motion has formed \citep{WAT:2009, HGS:2013, PAH:2014}. Here, the optical emission can be split into two components: i) continuum disk emission usually modelled with a multi-temperature, geometrically thin disk \citep{FAR:2002}, and ii) orbitally modulated optical emission from the companion's pulsar-wind-heated surface and ellipsoidal modulations \citep{Papitto:2013}. 

Recently, the optical and near-IR emission from \psr\ has been probed in extensive detail using photometry, spectroscopic  \citep{TA:2005,KCV:2018,  SDG:2018} and polarimetric observations \citep{BDC:2016, HK:2018}. The fast photometric observations have revealed evidence for  mode switching as observed previously in X-rays \citep{SLN:2015,PRC:2018}. Secondly, compared to the X-ray light curves, a significantly higher portion of the optical emission is observed in `flare' mode \citep[$\sim 15.6-22\%$;][]{PRC:2018,KCV:2018}. Interestingly, a recent work by \citet{KCV:2018} shows that the fluence of a flare depends on the wait time since the previous flare. Lastly, polarimetric observations \citep{BDC:2016} reveal linear polarization in this state. \citet{HK:2018} show that the polarization angle
diminishes during flares and attribute this to the rise of another polarization component during flares. Finally, \citet{PAS:2019} showed that optical emission is correlated with X-rays during the flare mode.  

Moreover, our past observations of \psr\ in the LMXB state revealed a complex UV spectrum dominated by the continuum of a truncated accretion disk far from the light cylinder of the neutron star. In addition, the spectrum presents a diverse range of strong and broad emission lines (${\sim}400$--$1000$ \kms) and many different ionization species, suggestive of an outflow from the system \citep[][Hern\'andez Santisteban et al., in prep.]{Her:2016}. These peculiar UV properties (ionization structure and kinematics) can be explained as being formed in a magnetically driven outflow, similar to the one observed in the accreting white dwarf system AE Aqr \citep{Eracleous:1996aa}. 
In addition, the near-UV (and possibly the far-UV) light curves show variability and hints of bi-modality, similar to those present in the X-rays \citep{ABP:2015, BAB:2015}. However, the lack of simultaneous X-ray data prevented us from reaching any definitive conclusion regarding their correlated nature. 

\subsection{Radio emission}
In the RMSP state, the tMSPs show coherent radio pulsations at the spin period of the neutron star. In addition, radio eclipses are seen due to clumpy, intra-binary ionized material \citep[e.g.,][]{AKH:2013}. 

During the LMXB state, the radio pulsations disappear and, instead, variable, roughly flat-spectrum continuum radio emission, suggestive of a collimated outflow, is observed \citep{DMM:2015}. Moreover, using coordinated Very Large Array (VLA) and \textit{Chandra} observations, \citet{BDM:2018} uncovered an anti-correlation between radio and X-ray brightness of \psr: radio brightness consistently peaks during the X-ray low modes. The radio light curve is observed to vary on timescales of minutes after mode switching, while X-ray mode switches occur on timescales of seconds followed by stability in other wavebands, supporting the idea that the continuum radio emission we observe is coming from regions at some distance from the neutron star compared to other wavebands.

\subsection{Pulsations at different wavelengths}

In the LMXB state, the pulsed radio emission from tMSPs disappears but \citet{ABP:2015} and \citet{JAH:2016} have demonstrated the presence of X-ray pulsations at the neutron star spin period, which are present only in the X-ray high mode. In fast photometry observations, \citet{APS:2017} have found optical pulsations during the high mode in the LMXB state. In addition, \citet{PAS:2019} identify optical pulsations (albeit with reduced flux) during flares.  Using the X-ray pulsations, we recently showed that \psr's spin-down is enhanced in the LMXB state \citep[][Jaodand et al.~\textit{in prep.}]{JAH:2016}. We suggested that the enhanced negative torques on the neutron star could be contributed by accretion material being ejected through a mechanism like propeller-mode accretion \citep{PT:2015}, while the pulsar wind remains active. While \citet{ABP:2015} argued that X-ray pulsations arise from channeled accretion of the material onto the neutron star surface, \citet{APS:2017} offer an alternate explanation in which synchrotron emission from the active pulsar wind shocking the accretion material surrounding it produces the optical photons. \citet{PAS:2019} show that the optical pulsations remain active during the flares and simultaneous optical and X-ray timing observations, and argue that regions for X-ray and optical emission should lie within ($3\times(sin~\iota))^{-1}$ km or a few kms of each other. They also extend the flux power law from X-ray emission to optical to show that the pulsations must arise in the same region. In this work, we show a slightly different conclusion.


\subsection{Multi-wavelength campaigns}
The wealth of observational information from tMSPs in the RMSP and LMXB states enables us to weave together various multi-wavelength observations in order to construct a concrete picture of the long-lived, sub-luminous, accretion-dominated LMXB state in the tMSPs. 

Previously, these key campaigns have compared joint variability with some combination of X-ray, optical, NIR and/or radio observations such as in \citet{deMartino:2013, TLK:2014, BPH:2014, BDM:2018, SDG:2018, Coti:2018, PAS:2019}.  For instance, \citet{BPH:2014} showed that intense X-ray flares are accompanied by simultaneous UV/optical flaring.  
Furthermore, the \citet{Coti:2018} campaign with X-ray Multi-Mirror (\xmm) and Nuclear Spectroscopic Telescope Array (\nustar) observations showed that the soft and hard X-rays are perfectly correlated with no lags. Moreover, \citet{SLN:2015} showed the presence of possible joint X-ray, UV, and optical mode switching. These results were ascribed to the presence of a hot, clumpy accretion flow in the inner edges of the accretion disk, which can explain the peculiar top-hat light curves in X-rays and optical. Later, in \citet{SDG:2018}, this argument was extended with joint optical and near-IR observations to the near-IR component believed to arise from plasmoids in the hot accretion flow and reprocessing of the optical emission. In \citet{PAS:2019} a strictly simultaneous timing and variability campaign is presented for X-ray and optical emission and  precise time lags are observed between the pulsations indicating a similar region of origin. 

These previous campaigns have highlighted the power of a simultaneous multi-wavelength approach. \textit{However, a simultaneous global campaign spanning high-time-resolution optical, UV, and soft and hard X-ray emission spanning multiple binary orbits had not yet been performed}. Such a campaign is critical to understand if both optical and UV emission originate from reprocessing of X-ray emission or changes at outer edges of the accretion disk. Finding time lags or leads between X-ray and UV emission can inform the mechanisms driving the persistent, low-level accretion regime in tMSPs. A deeper understanding of the optical and UV (if present) pulsed emission similar to the X-ray pulsations is also called for. 

Motivated by previous multi-wavelength results and remaining open questions, we obtained simultaneous observations with \xmm, \nustar, Hubble Space Telescope (\textit{HST}) and \kep\ targeting \psr. A summary of these observations is presented is \S2. The data analysis undertaken to probe joint variability
and pulsation searches is presented in \S3. In \S4, we present our results. The implications of our findings are discussed in detail in \S5.

\section{Observations and data analysis}
\label{subsec:obs}
\subsection{Observation and data reduction}
\label{subsec:tables}
Here we describe the observations and data reduction process for each observatory used in our campaign.  Table~\ref{tab:obssummary} summarizes these observations.


\subsection{\xmm} 
The \textit{XMM-Newton}  observations of PSR J1023+0038 considered here were obtained on 2017 June 13 (ObsID 0803620501). The European Photon Imaging Camera (EPIC) pn instrument \citep{SBD:2001} was operated in ``Timing'' mode, which offers a time resolution of 29\,$\mu$s by foregoing one imaging dimension. The two EPIC MOS cameras \citep{TAB:2001} were employed in ``Small Window'' mode to minimize the effect of pileup.

The X-ray event data were reduced and extracted using the Science Analysis Software (SAS\footnote{The \textit{XMM-Newton} SAS is developed and maintained by the Science Operations Centre at the European Space Astronomy Centre and the Survey Science Centre at the University of Leicester.}) package version \texttt{xmmsas\_20180620\_1732-17.0.0}. The data were inspected for intervals of strong background flaring and none were found. The source photons were extracted from the pn Timing mode observations from a region
of width 6.5 pixels in the imaging (RAWX) direction centered
on row 37. This corresponds to an angular size of 27" in the RAWX detector coordinate, which
encircles $\sim$87\% of the energy from the point source at 1.5 keV. The MOS1/2 source events were extracted from circular regions of radius 36$''$, which encircle $\sim$88\% of the total point source energy at $\sim$1.5 keV.
Only events in the 0.3--10 keV range were considered and were further filtered using the recommended flag and pattern values. The resulting exposures were 22.7 ks for MOS1, 22.6 ks for MOS2, and  20.7 ks for the pn. 

To produce the time series X-ray light curve, the data were grouped in 10 second intervals, and each bin was background subtracted and exposure corrected using the \texttt{epiclccor} tool in SAS. The resulting light curves from the three detectors were then stacked to produce a total 0.3-10 keV \textit{XMM-Newton} EPIC light curve.  The photon arrival times were  barycentered using the SAS task \texttt{barycen} assuming the DE405 solar system ephemeris and the pulsar's astrometric ephemeris presented in \citet{DAB:2012}.

The \xmm\ Optical Monitor \citep{MBM:2001} was used with the $B$ filter, which has a band pass of  $3800-5000$ \AA\ centered on $4392$ \AA. The operating mode was set to ``Image Fast'' to provide fast photometry. The  photometric $B$ filter OM data were produced using the SAS \caps{omfchain} pipeline tool using the default set of parameters and a 10 s time binning.

\subsection{\hst} 
NUV observations were taken with the Space Telescope Imaging Spectrograph (STIS) on-board \hst\ in two visits. The first visit consisted of 3~orbits (orbits a-c hereafter), for a total exposure time of 8.1 ks, under the DDT program 13630. The second 4~orbits --- which were observed in the multiwavelength campaign --- (orbits 1-4 hereafter), consisted of 8.7 ks total exposure time, under the DDT program 14934. All observations were taken in TIME-TAG mode with a time resolution of 125 $\mu$s with the MAMA detector. We used the G230L grating and a 52$\times$0.2 slit to achieve a spectral resolution of $R\sim500$ in the range of $1570-3180$\AA. The spectra were reduced and extracted through the standard \caps{calstis} pipeline within \caps{iraf/stsdas}. The event files were barycentered using the \caps{odelaytime} routine. We further processed the event files to assign individual wavelengths with a custom routine\footnote{\url{https://github.com/Alymantara/stis_photons}} and select those events between 1800 and 2800 \AA\ in order to avoid any emission lines. In addition, we selected events within 10 pixels of the center of the slit to minimize contributions from the rest of the detector.
We note that while the relative time-stamp accuracy of photons with STIS within an observation is accurate to the nominal value, the absolute timing can differ between $\sim1-2$ s (\hst\ helpdesk priv. comm.). Therefore, any relative phasing differences between \hst\ and other observatories in timescales lower than $\lesssim2$ s are likely instrumental. We show this explicitly in the search for UV pulsations in Section~\ref{sec:pulse}.

\begin{deluxetable*}{lllllll}
  \caption{\label{tab:obssummary}Log of observations}
 \tablehead{
 \colhead{Facility} & \colhead{Wavelength (nm)} &\colhead{Orbit}& \colhead{Start BMJD} & \colhead{End BMJD}& Exposure& \colhead{Comments}  \\
}
 \startdata
 \hst       & 180--280   & a     & 56774.6148329384 &	56774.6420063143    &  2.3 ks  &   \\
            & 180--280   & b     & 56774.673056799 &	56774.7075329955    &  2.9 ks  &  \\
            & 180--280   & c     & 56774.739428658 &	56774.7739048626    &  2.9 ks &  \\
\hst       & 180--280   & 1     & 57917.6342879827 &   57917.6611050226    &  2.3 ks  & Mode switch  \\
            & 180--280   & 2     & 57917.7005084648 &   57917.7283153673    &  2.4 ks  & Mode switch \\
            & 180--280   & 3     & 57917.7667407978 &   57917.7905611494    &  2.1 ks & Flare \\
            & 180--280   & 4     & 57917.8329616436 &   57917.8556557611    &  2.0 ks& Flare \\
 \nustar    & 0.02--0.41 & 1     & 57917.6342879827 &   57918.1515853949    &  44.7 ks &  \\
  \xmm      & 0.12--4.13 & 1     & 57917.6342879827 &   57917.9771643519    &  29.6 ks &  \\
  \kep      & 400--800   & 1     & 57917.6342879827 &   57917.8694315006    &  20.3 ks &  \\
 \enddata

 \end{deluxetable*}

\begin{figure*}
\plotone{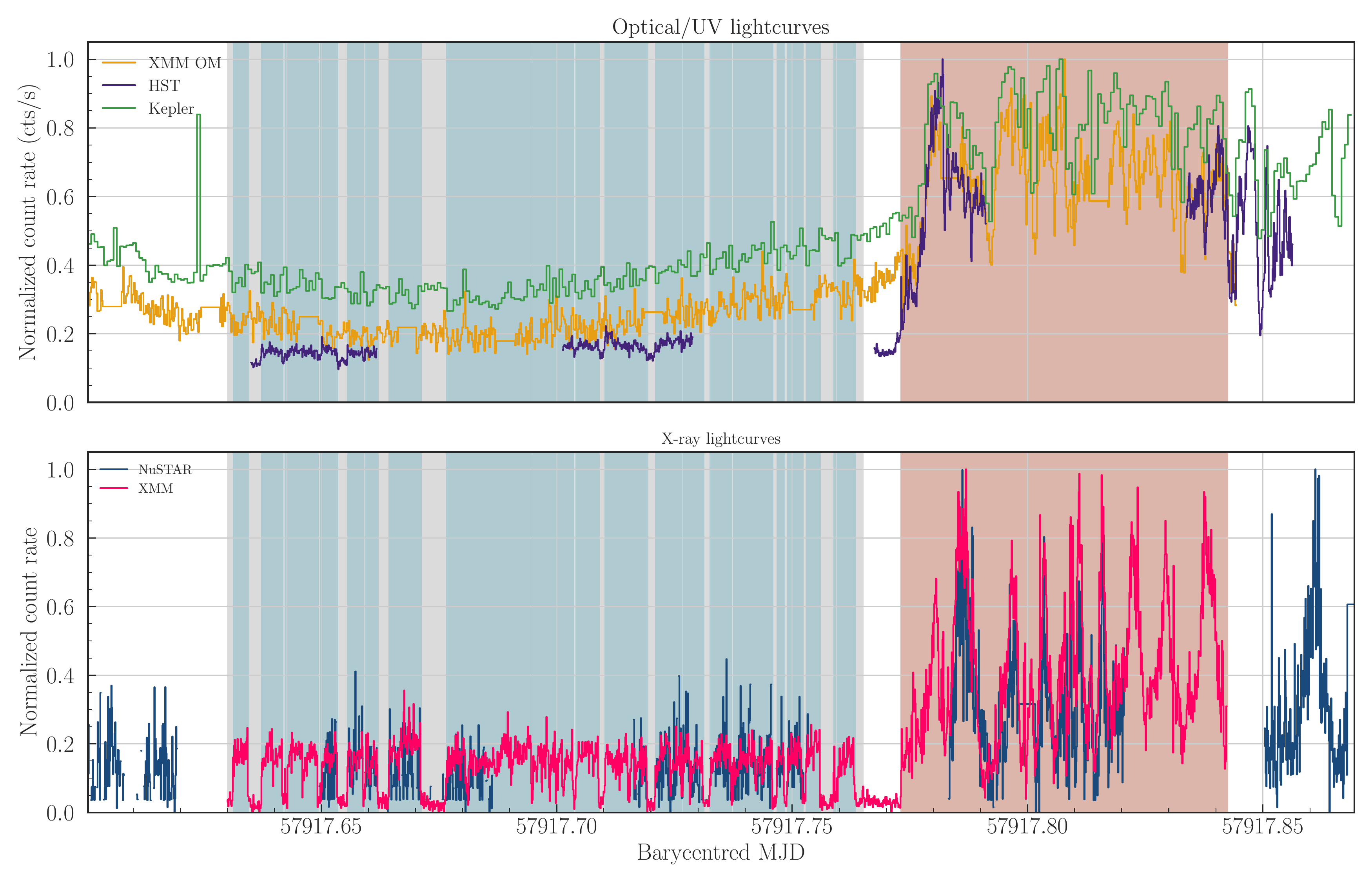}
\caption{Multi-wavelength light curves of \psr\ during the observing campaign presented here.  For each instrument, the count rate is normalized separately.  \textit{Top panel:} optical and UV light curves from \kep\ ($400-800$\,nm), \textit{XMM-Newton} OM (B filter; $390-490$\,nm) and \textit{HST} ($180-280$\,nm).  \textit{Bottom panel:} X-ray light curves from \textit{XMM-Newton} ($0.3-10$\,keV) and \nustar\ ($3-79$\,keV).  The vertical shaded regions indicate X-ray high (blue), low (grey) and flare (red) modes, as derived from the \textit{XMM-Newton} light curve.  Unclassified spans are left with a white background; these are either regions with no \textit{XMM-Newton} coverage, or where we chose not to classify because of poor overlap with \textit{HST}.
\label{fig:lc_mw}}
\end{figure*}

\subsection{\nustar}
The \textit{Nuclear Spectroscopic Telescope Array} (\nustar; \citealt{2013ApJ...770..103H}) is the first focusing hard X-ray telescope. It consists of a pair of co-aligned telescopes with grazing incident optics to focus X-rays onto Focal Plane Modules (FPMA and FPMB) that record photons with energies between 3 and 79 keV.

Following a DDT request, \psr\ was observed by \nustar\ for 55 ks during a time stretch starting on 2017 June 13 at 12:06 UT and ending on June 14 at 04:01 UT. We used \texttt{nuproducts}, incorporated in HEASOFT (v. 6.23), to extract data products. A circular region with a radius of 60$''$ was used to extract source events, whereas a region of similar dimensions, placed on the same chip away from the source, was used to extract background events. 

Initial inspection of the FMPA/FMPB X-ray spectrum, combined using \textsc{addascaspec}, showed that \psr\ was detected above the background up to an energy of $\sim 30$~keV. Light curves, with different time bins, were therefore extracted using an energy range of 3--30~keV. The \nustar\ light curves were barycentered using \texttt{barycorr}, using the most up-to-date clock file at the time of our observations (nuCclock20100101v072.fits) and the astrometric ephemeris of \psr\ from \citet{DAB:2012}. 

\subsection{\kep}  
\kep/\textit{K2} observed \psr\ between 2017 June 01 05:21:11 UT and 2017 August 19 21:56:19 UT, as part of mission 14. Using Mikulski Archive for Space Telescope (MAST) archive, we obtained the raw-format, target pixel, short-cadence (cadence of 58.8 s) observational data. The target point-spread function was observed to be asymmetric as \psr\ was in module 16.4 which is one of the outermost modules. However, the module should be unaffected by Moire Pattern Drift noise. On account of this extended point-spread-function of \psr\, and target displacement due to small spacecraft jitters we created a large target mask. The use of large target mask was justified as no other sources are present in the vicinity of \psr. Moreover, we also created a a background mask. After this masking operation we obtained a source lightcurve for \psr. The same light curve used here has also been used in \citet{PRC:2018}, where the employed \textit{K2} mask is shown in their Figure~2, as well as the broad-band power spectral density for \psr.



\section{Data Analysis}
\label{subsec:da}
\subsection{Timestamp verification and mode classification}

After having performed the data reduction for the various telescopes as outlined above, we obtained a list of valid photons with their arrival times in modified julian date (MJD) or mission elapsed time, and energy associated with each photon. Each of the telescopes assigns photon timestamps in a differing manner. In order to ensure the uniformity of time conventions we chose to convert timestamps in all the observation to MJD format with barycentric dynamical time (TDB) scale. 

\begin{figure*}
\begin{center}
\includegraphics[width=0.75\textwidth]{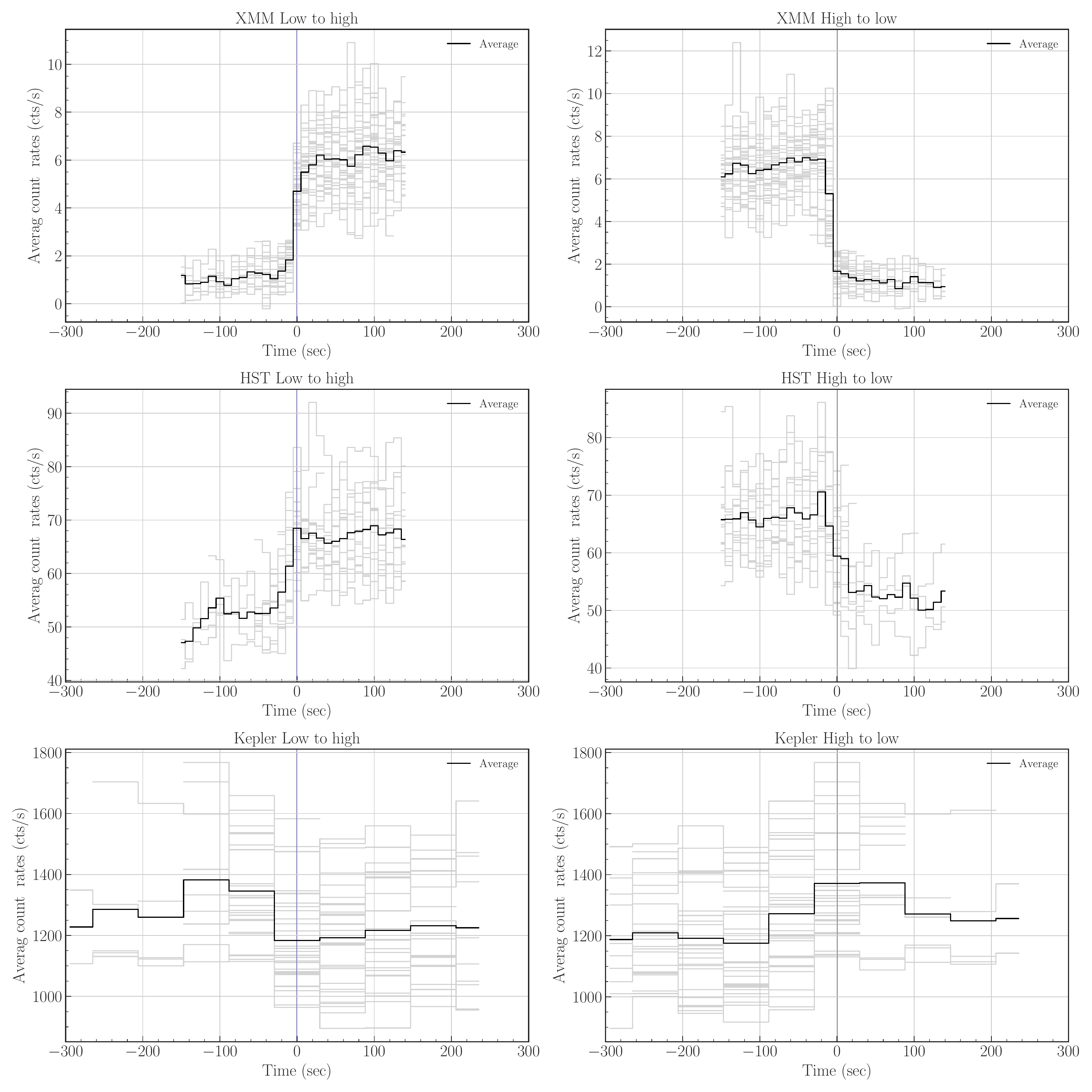}
\caption{Stacked mode plots showing how the count rate changes (on average) during the low-to-high (left column) and high-to-low (right column) transition for \textit{XMM-Newton} (top; $0.3-10$\,keV), \textit{HST} (middle; $180-280$\,nm) and \kep\ (bottom; $400-800$\,nm).  The modal classification and transition is based on the \textit{XMM-Newton} light curve and applied to all simultaneous data from \textit{HST} and \kep.  In each panel, the light grey bins show the count rates from individual segments of the light curves; the black bins are a weighted average of these.  These are shown with respect to the time before or after a mode transition, which is also marked with a vertical blue line.  Note that when \textit{XMM-Newton} and \textit{HST} are in a low mode, \kep\ appears to be in a high mode, and vice versa.
\label{fig:mode_switches}}
\end{center}
\end{figure*}

To obtain the light curves presented in Fig.\ref{fig:lc_mw} we binned the \hst\ and \nustar\ datasets at $10$\,s. Here the bin edges, and observation start and end time limits were dictated by the \xmm\ observation. The \kep\ dataset is received in light curve form, with bin sizes set by the exposure and readout time limit to $59$\,s. Although Fig.~\ref{fig:lc_mw} and Fig.\ref{fig:mode_switches} use \kep\ dataset with $59$\,s binning, the start and end times for this light curve are also obtained from \xmm. For all other figures, wherever we have used the \kep\ dataset we re-binned the \xmm, \hst\ and \nustar\ datasets according to \kep\ bin edges and while still setting the time limits with \xmm.  

We used only the \xmm\ light curve to classify times into low, high, or flare modes. Fig.~\ref{fig:lc_mw} shows that the flares are confined to the end of the \xmm\ observation. We therefore classified all times after MJD~57917.773 as flare-mode, and all times between MJD~57917.765 and MJD~57917.773 as ambiguous. For the remaining times (before MJD~57917.765), we used the exposure-corrected \xmm\ light curve to create a histogram of the count rates in a manner similar to Fig. A1 in \citet{ABP:2015}. Using this histogram, we chose a count rate limit of 2.67\,s$^{-1}$ 
to classify the observed count rates into low and high modes. We group these times into a list of low, high, and flare intervals. These \xmm\ based time classifications are used as mode markers for \hst, \kep\ and \nustar\ observations. These markers serve as the basis for the modes indicated in various plots in this work: e.g., the varying background colors corresponding to a specific mode in Fig. \ref{fig:lc_mw}. 

\subsubsection{Mode stacking}
We use the above mode classification to identify transition edges --- to the nearest light curve bin --- from low-to-high and high-to-low modes. For every transition edge we gather data from the `before' and `after' modes, up to a maximum length of 300~s in each direction. Next, we stack these data-chunks around transition edge, averaging all bins that are a common time before or after a mode transition (Fig. \ref{fig:mode_switches}). 

\subsection{Auto and cross correlation analysis}

To compute the cross-correlation between two time series $S_1(t)$ and $S_2(t)$ (for autocorrelation we choose $S_1=S_2$), we use the auxiliary time series $V_1(t)$ and $V_2(t)$, which are one where there is valid flare-mode data at time $t$ in the corresponding time series, and zero elsewhere (including past the end of each time series). For each delay $\tau$ we define the usable data domain $D(\tau) = \{t : V_1(t)V_2(t+\tau)\neq 0\}$, that is, the values of $t$ where both (shifted) time series are valid. We then compute the correlation function as
\[
{\mathrm{Corr}}(\tau) = \frac{\sum_{t\in D(\tau)}S_1(t)S_2(t+\tau)}{\sqrt{\sum_{t\in D(\tau)} S_1^2(t)}\sqrt{\sum_{t\in D(\tau)} S_2^2(t)}},
\]
where $\tau$ is the lag between the two series. This computation has the property that correlating a series with a positively scaled copy of itself gives $\mathrm{Corr}(\tau)=1$ at the correct lag regardless of where either series is valid (as long as there is non-zero overlap).

\subsubsection{Pulsation Search}

\begin{figure*}
\plotone{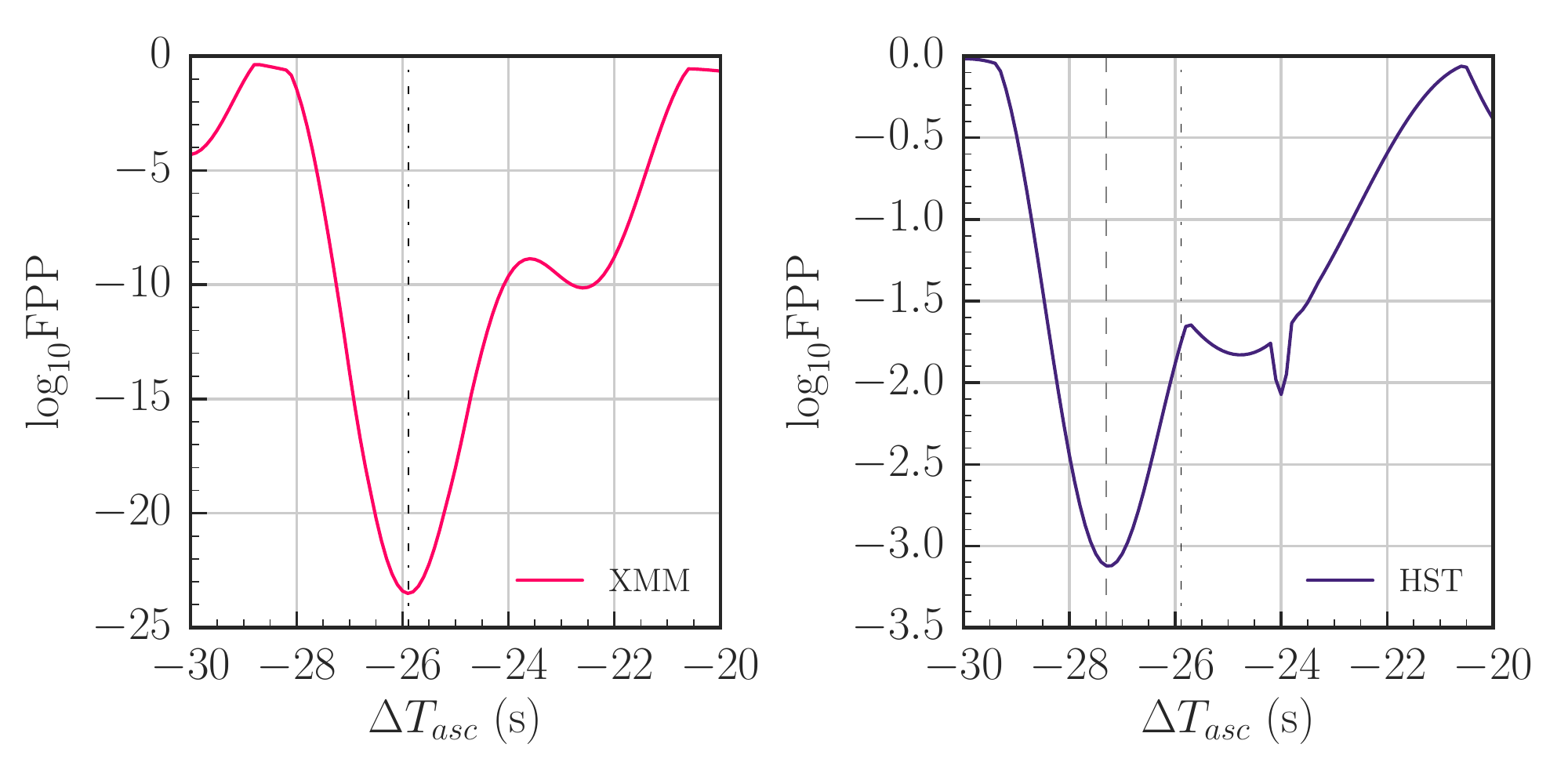}
\caption{\textit{H} scores, expressed as false-positive probabilities (FPP), as a function of the time of ascending node $T_{\rm asc}$ compared to a reference orbital ephemeris.  \textit{Left:} \textit{XMM-Newton} shows a clear optimization at T$_{asc,xmm} = -25.9$\,s (marked by the dot-dashed vertical line on both panels).  \textit{Right:} The corresponding optimization for \hst\ where, because the absolute timing is only accurate to within $\sim 1-2$\,s, the best-fit value deviates from that found with \textit{XMM-Newton}. This value  T$_{asc,hst} = -27.3$\,s is denoted by the dashed grey line. 
\label{fig:tasc_search}} 
\end{figure*}

To conduct the UV pulsation search analysis we roughly follow the same recipe as outlined in $\S$2.5,$\S$2.6 and $\S$3.1 of \citet{JAH:2016}. We start with the X-ray ephemeris presented in \citet{JAH:2016}, which has been extended to include a new set of \xmm\ observations obtained in 2017 (Jaodand et al., \textit{in prep}). We use this ephemeris and fit for the time of ascending node ($T_{asc}$)
to model the orbit in the \xmm\ observation. Note that this analysis is only conducted on data corresponding to times when \psr\ is in high mode (as determined from the simultaneous X-ray data). Because of the uncertainty in the absolute timing of the \hst\ data, we searched a range of three seconds around the \xmm-derived (true) $T_{asc}$ value to find a $T_{asc}$ value for the \hst\ data that compensates for the clock offset. The clock offset (as indicated by the dip in Fig. \ref{fig:tasc_search}) for \hst\ at the time of this observation\footnote{We do not expect this offset to change significantly between Hubble orbits.} appears to have been $-1.4$~s. We cannot determine this quantity to the sub-millisecond accuracy necessary to obtain a meaningful pulse phase offset between the \hst\ and \xmm\ data.

The ephemerides obtained for \xmm\ and \hst\ with adjusted $T_{asc}$ values (T$_{asc,xmm} = -25.9$\,s, T$_{asc,hst} = -27.3$\,s) are then used to fold the respective datasets. In this way, we obtain the pulse profiles shown in Fig.\ref{fig:pulse_profs}. More detailed information on the LMXB state X-ray-derived ephemeris, the precise folding procedure and the H-test-based significance testing is given in \cite{JAH:2016}.




\section{Results}

The simultaneity between instruments spanning from 1.55\,eV to 79\,keV (800\,nm to 0.015\,nm) has allowed us to detect correlated changes in the broadband behavior of \psr.  Here we discuss the observed phenomena related to low/high modes, flaring and pulsations.  Though the \textit{XMM-Newton} OM provided contemporaneous data (Fig.~\ref{fig:lc_mw}), these data are of lower sensitivity and largely redundant with the wavelength coverage provided by \kep; hence, we do not discuss them in detail below.  Note that brightness variation in the optical bands is also affected by the ellipsoidal modulation of the companion star (the sinusoidal modulation visible in the \kep\ and \textit{XMM-Newton} OM light curves shown in Fig.~\ref{fig:lc_mw}).  However, this is on significantly longer timescales compared to the few-minute timescale moding/flaring and millisecond pulsations that are the main focus of this present work.

\subsection{Low and high modes}

\begin{figure*}
\plotone{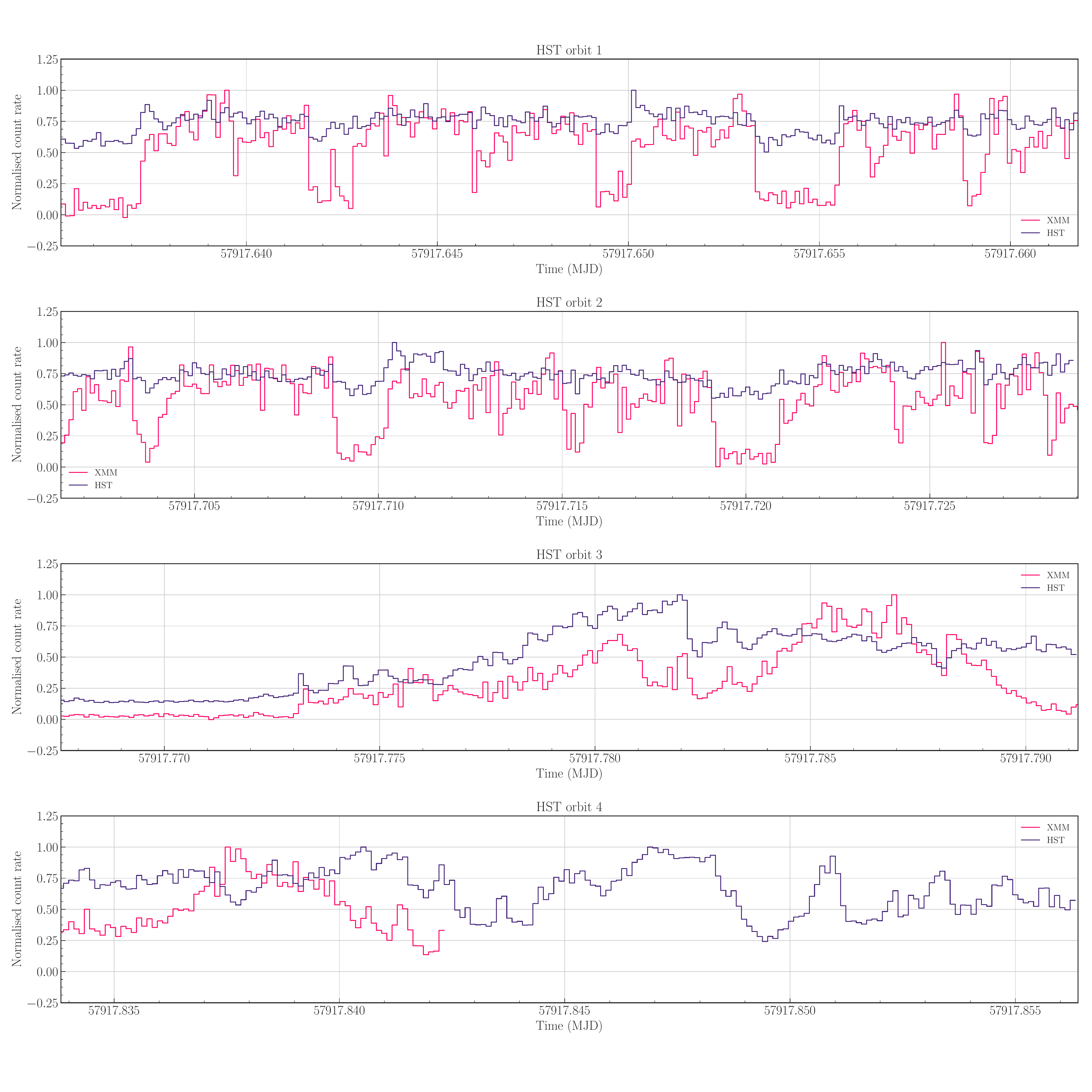}
\caption{Multi-wavelength light curve of \psr\ as observed by \textit{HST} ($180-280$\,nm) and \textit{XMM-Newton} ($0.3-10$\,keV). Here we compare these light curves for each \textit{HST} orbit separately. Orbits 1 and 2 show high and low moding behavior, whereas orbits 3 and 4 are dominated by flaring.}  
\label{fig:lc_hst}
\end{figure*}

\begin{figure*}
\begin{center}
\includegraphics[width=0.7\textwidth]{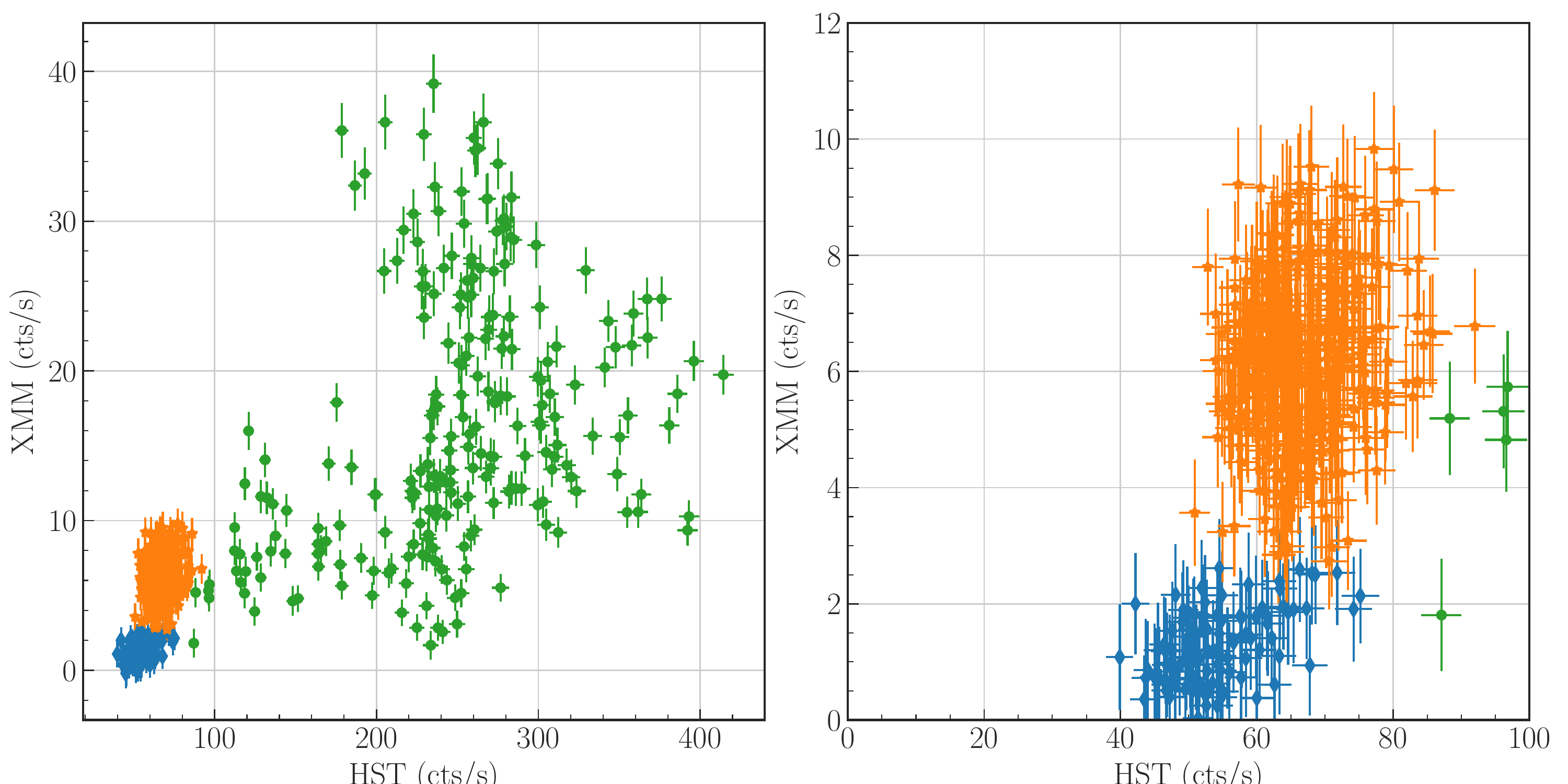}
\caption{Comparison of strictly simultaneous \textit{XMM-Newton} ($0.3-10$\,keV) and \textit{HST} ($180-280$\,nm) count rates.  Each point is from an individual 10-s bin.  The colors represent the low (blue), high (orange) and flare (green) modes, as derived from classifying the \textit{XMM-Newton} light curve.  \textit{Left:} the full count range; \textit{Right:} zoom-in on the low and high modes.  Though the flares produce a wide range of count rates, the low and high modes show a more reproducible behavior for both \textit{XMM-Newton} and \textit{HST}.
\label{fig:xmm_hst_counts}}
\end{center}
\end{figure*}
The well-established switches between X-ray low and high modes \citep[e.g.][]{BAB:2015} are clearly detected in the \textit{XMM-Newton} ($0.3-10$\,keV) light curve (Fig.~\ref{fig:lc_mw} and Fig.~\ref{fig:lc_hst}) and appear to extend through to the hard X-ray band probed by \nustar\ ($3-79$\,keV).  This has previously been established by \citet{TYK:2014} and \citet{Coti:2018}, and thus we do not investigate it in more detail here.

In the first two \hst\ orbits, where there is no flaring (Fig.~\ref{fig:lc_hst}), we find clear evidence for mode switching in UV ($180-280$\,nm; Fig.~\ref{fig:lc_mw}) occurring synchronously with the X-ray mode switches (Fig.~\ref{fig:mode_switches}).  The mode switching is less obvious in \hst\ data compared to the X-ray data (Fig.~\ref{fig:xmm_hst_counts}): there is only a $\sim25$\% average increase in count rate from low to the high mode, compared to the $\sim600$\% average increase seen in \textit{XMM-Newton}.  The non-moding component in UV is further discussed in \S\ref{sec:disc}.

 The simultaneity with X-ray data has also allowed us to identify subtle mode switching in the optical to NIR band ($400-800$\,nm) provided by \kep\ (Fig.~\ref{fig:mode_switches}).  The mode switches appear to be simultaneous with the X-ray transitions; however, the low time resolution of the \kep\ data complicates this comparison and reduces the sensitivity to short lags between the two bands. Most importantly, we find that the \textit{\kep\ optical to NIR signal changes in the opposite direction to the X-ray and UV}: when an (X-ray) low mode begins, the optical/NIR increases by ${\sim}10\%$ and vice versa (Fig.~\ref{fig:mode_switches}).  Because the X-ray data provides the most well-established and most easily observed mode switches, we continue to call periods when the X-rays are steady and bright ``high modes'' and those when the X-rays are steady but faint ``low modes''.

\subsection{Flares}

\begin{figure*}
\plotone{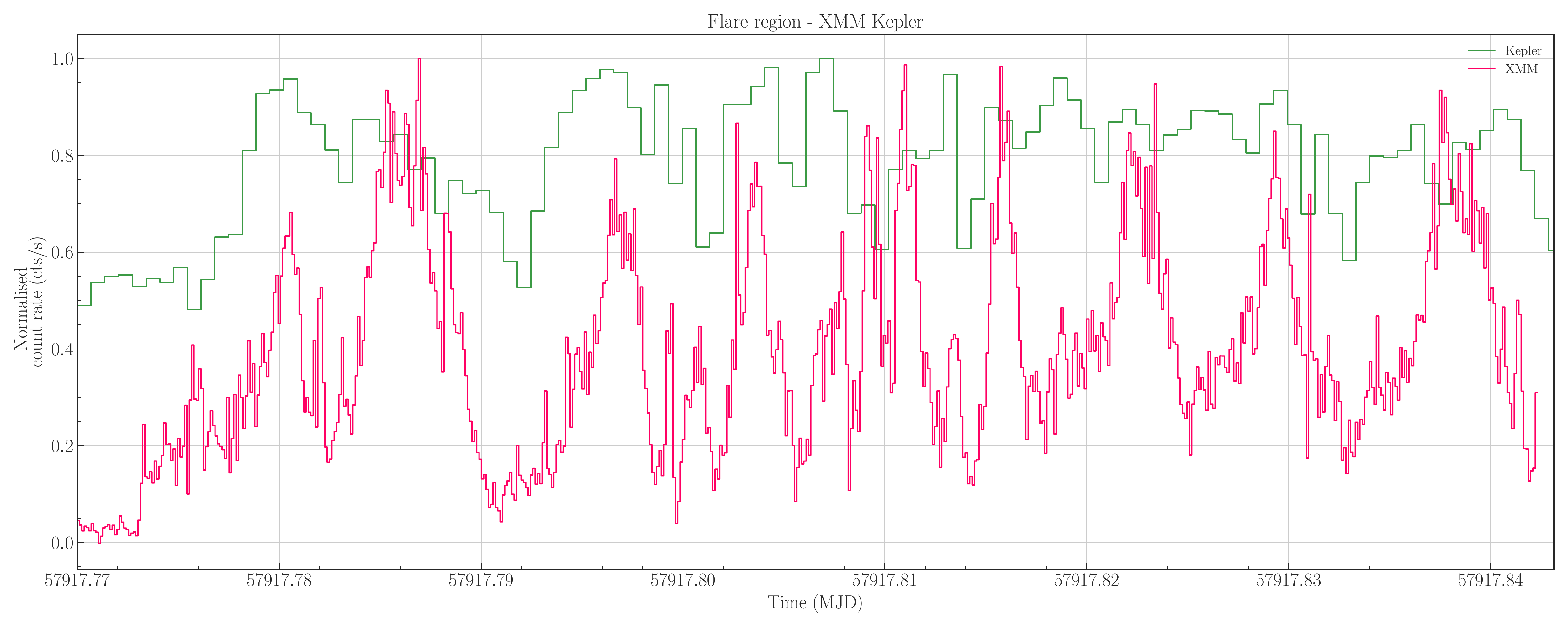}
\caption{Comparison of X-ray and optical light curves of \psr\ during the flaring period, as observed by \textit{XMM-Newton} ($0.3-10$\,keV) and \kep\ ($400-800$\,nm).
\label{fig:lc_flare}}
\end{figure*}

\begin{figure}
\plotone{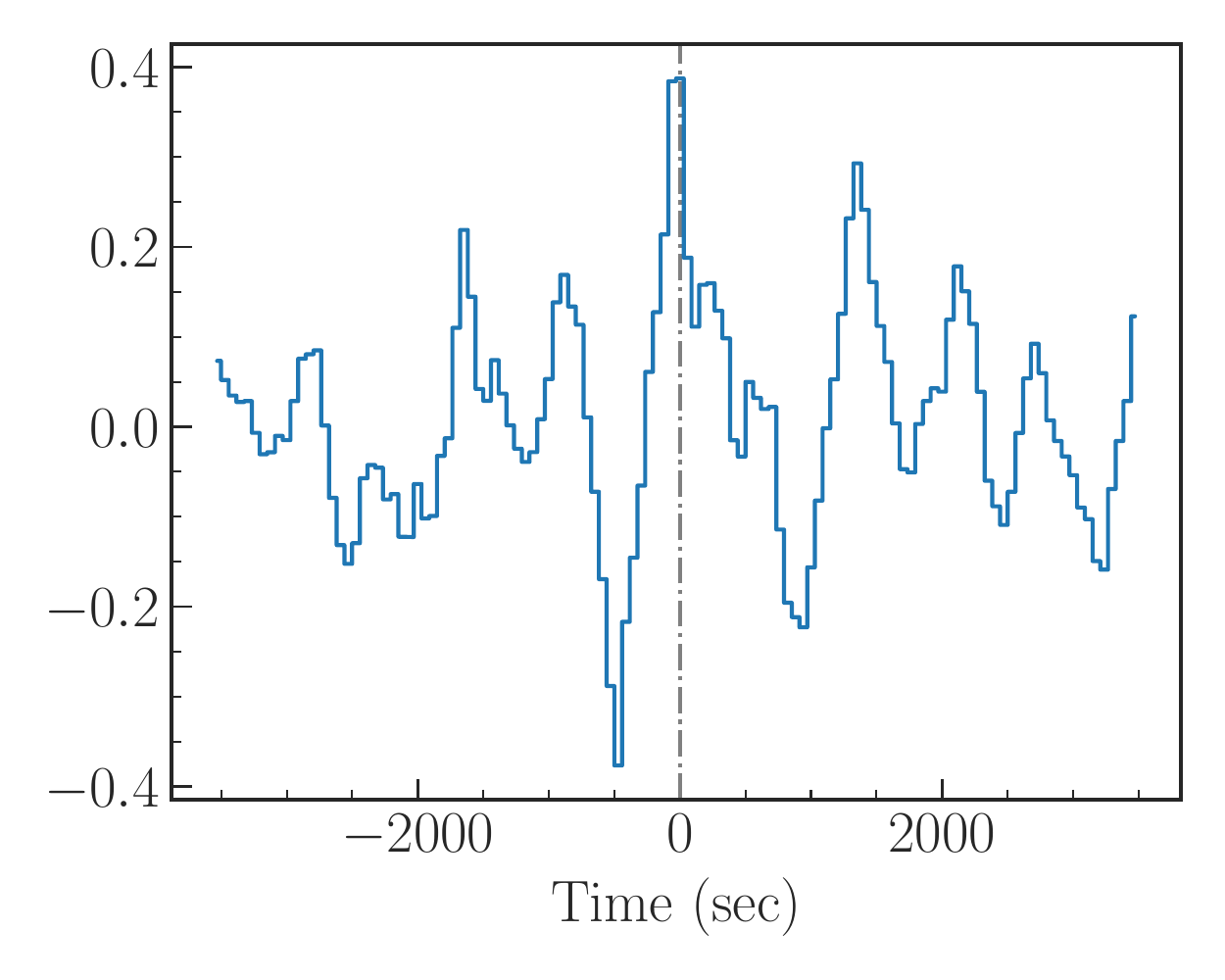}
\caption{Cross-correlation of the \textit{XMM-Newton} ($0.3-10$\,keV) and \kep\ ($400-800$\,nm) light curves during the flaring mode.  Before cross-correlation, the \textit{XMM-Newton} data was binned to the same $\sim59$\,s resolution as \kep.
\label{fig:cross_corr}}
\end{figure}

\begin{figure*}
\plotone{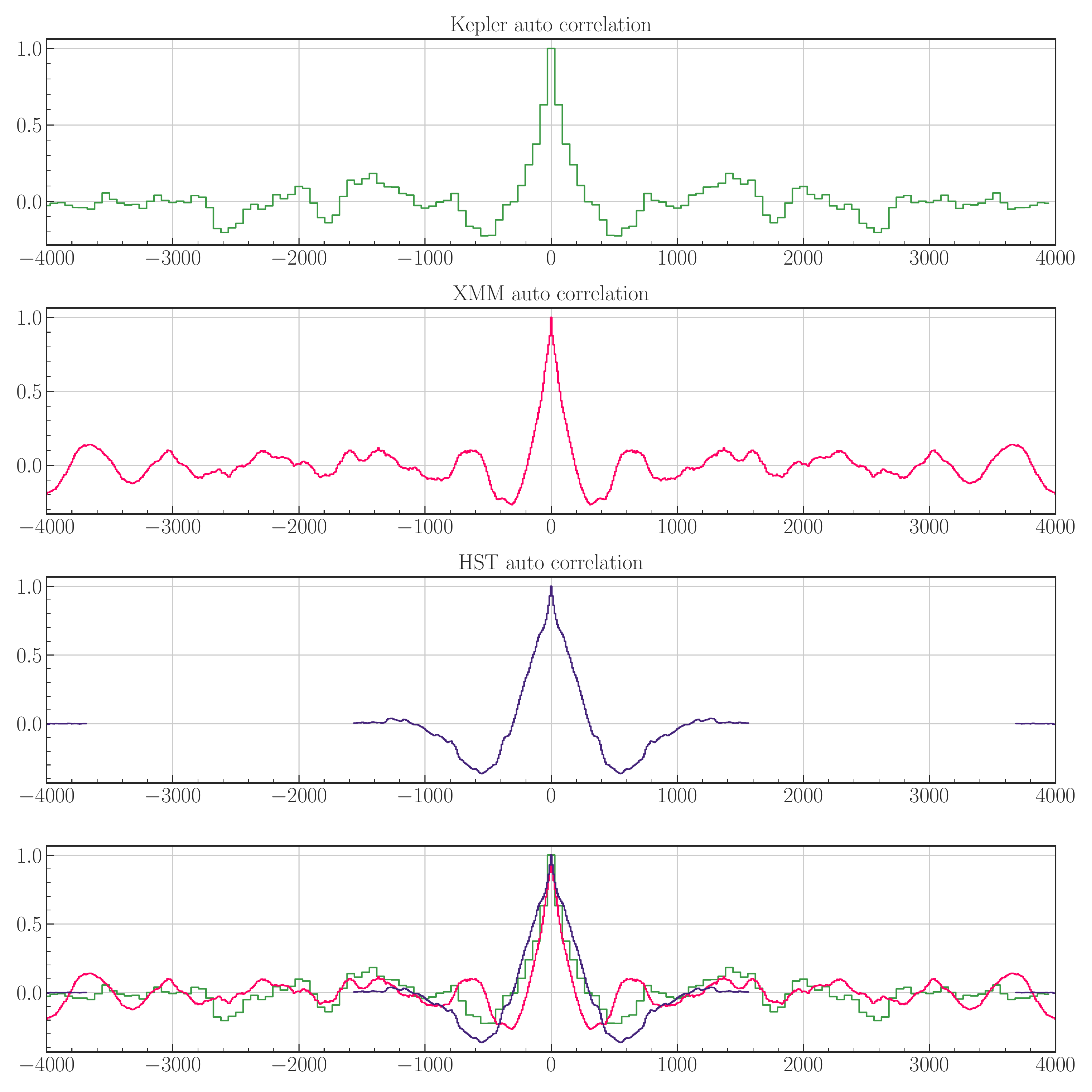}
\caption{Auto-correlations for \kep\ ($400-800$\,nm), \textit{HST} ($180-280$\,nm) and \textit{XMM-Newton} ($0.3-10$\,keV) light curves during the flare mode. The bottom panel is an overlay of all three auto-correlations, and demonstrates that the flare durations are similar.
\label{fig:auto_corrs}}
\end{figure*}

The observing campaign detected a flaring period lasting for at least $\sim2.5$\,hr (\psr\ still appeared to be flaring as observing ceased).  During this time, \textit{XMM-Newton} provides ${\sim} 1.7$\,hr of contiguous coverage; the only other instrument providing comparable coverage is \kep, though \hst\ and \nustar\ also clearly show flaring (Fig.~\ref{fig:lc_mw}).  The flaring period is characterized by quasi-periodic flares  (Fig.~\ref{fig:lc_flare}), \textit{such that the flares occur simultaneously in the optical, UV and X-ray bands (Fig.~\ref{fig:cross_corr}) and have similar durations of ${\sim}500$\,s on average (Fig.~\ref{fig:auto_corrs})}. On the shortest timescales, however, it is clear that no tight correlation between X-ray and UV flare brightness is present (Fig.~\ref{fig:xmm_hst_counts}).

\subsection{Pulsations}\label{sec:pulse}

\begin{figure}
\plotone{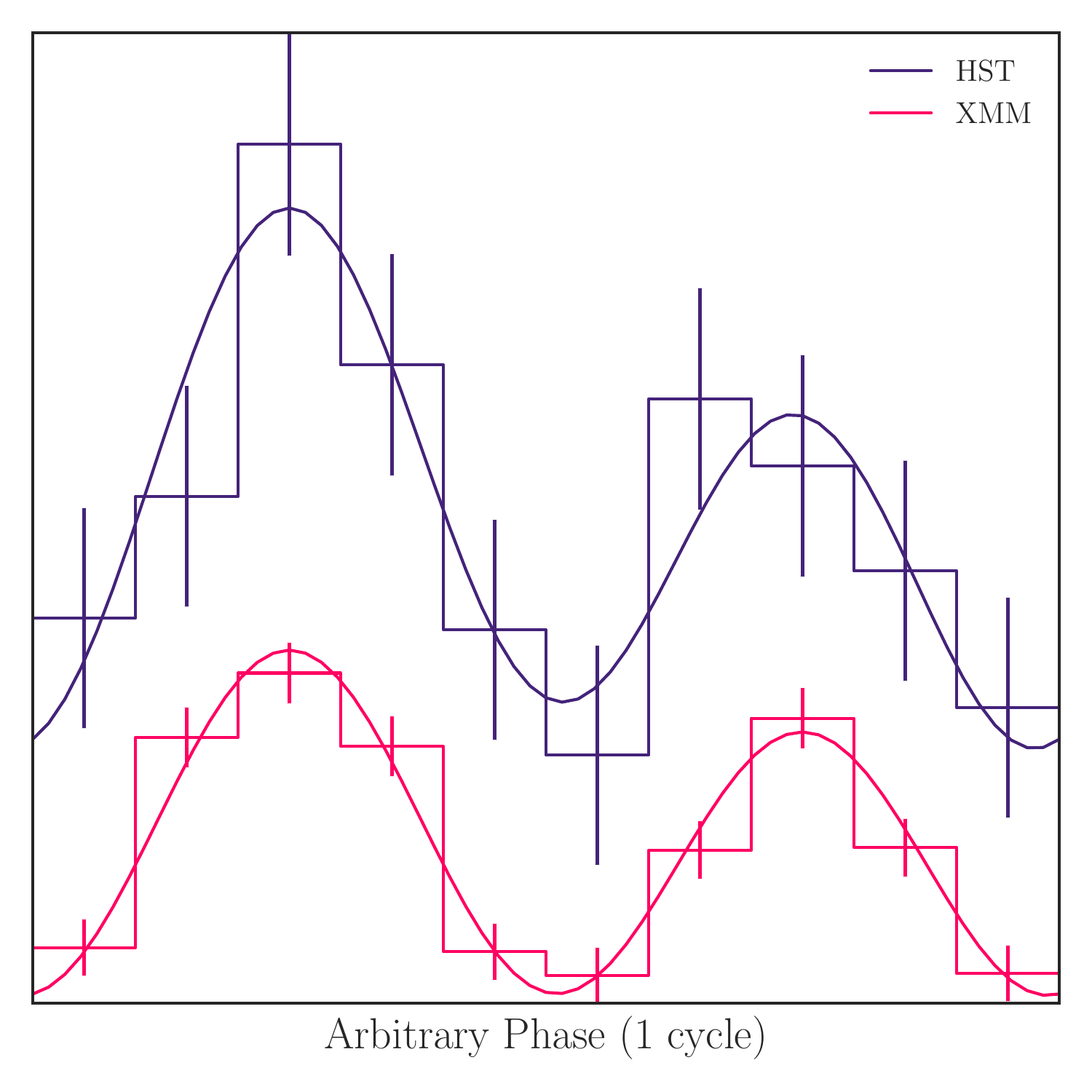}
\caption{Background-subtracted pulse profiles for \textit{XMM-Newton} and \hst\ observations folded using an extended timing solution based on \citet{JAH:2016} and adjusting for the local $\Delta T_{\rm asc}$ measurements.  An arbitrary vertical offset has been added to each profile so that they do not overlap. Pulsations are detected at 10.3$\sigma$ and 3.2$\sigma$ significance for \textit{XMM-Newton} and \hst, respectively.  Though they show a similar pulse profile structure, we caution that the absolute phasing is unknown because of the inaccuracy in the absolute time stamp of the \hst\ data.  The smooth curves plotted above the histograms are obtained directly from photon phases \citep[for more details, see][]{ABP:2015}.
\label{fig:pulse_profs}}
\end{figure}

After performing a search in $T_{\rm asc}$ to create a local orbital ephemeris and to correct for the large  uncertainty in the \hst\ absolute timing (Fig.~\ref{fig:tasc_search}), we detect pulsations in both the $0.3$--$10$\,keV X-ray and $180$--$280$\,nm UV bands (Fig.~\ref{fig:pulse_profs})\footnote{In principle, it may be possible to detect X-ray pulsations in the \nustar\ data as well, but this is complicated by clock variations as a function of the satellite's orbital phase \citep{TYK:2014}.}. The X-ray pulsations are detected with 10.3\,$\sigma$ (single-trial) significance and have a pulse fraction of $\sim 6$\%. The double-peaked pulse profile is consistent with those presented in \citet{ABP:2015} and \citet{JAH:2016}. The UV pulsations are detected with 3.2\,$\sigma$ (single-trial) significance and have a much lower pulse fraction of $0.82\pm0.19$\%.  Despite the lower formal significance of the UV pulsations, we consider this a clear detection because of the clear similarities in pulse morphology with the X-ray pulsations. 

Unfortunately, the absolute alignment between the two bands is unknown because of the \hst\ absolute timing accuracy. Given the limited statistics during the low mode, and the high background during the flares, it is not possible to constrain whether the UV pulsations turn on/off in sync with the moding --- as has been demonstrated for the X-ray pulsations \citep{ABP:2015}.


\section{Discussion}
\label{sec:disc}

Examining a power budget for the \psr\ system will shed some light on the relative importance of different physical processes. We know that power is available from three major sources: the spin-down of the neutron star   \citep[$7.12\times10^{34}$\,erg~$s^{-1}$][]{JAH:2016}, the companion's luminosity \citep[${\sim}1.4\times 10^{33}$\,erg~$s^{-1}$, ${\sim}2\%$ of the neutron-star spin-down;][]{SLR:2019}, and the accretion flow. We do not know the amount of mass transferred or its ultimate fate, so estimating the power available from accretion is very difficult. Nevertheless we can compare these power sources with the observed luminosities in different parts of the spectrum. The largest luminosity we observe is in the $\gamma$-ray ($0.1$--$300$ MeV), which is 8.4\%  of the spin-down power \citep[$6\times10^{33}$\,erg~s$^{-1}$][]{SAH:2014}. In X-rays ($0.3$--$10$ keV) we see different luminosities depending on mode: in the high mode 4.2\% ($3\times 10^{33}$\,erg~s$^{-1}$) and in the low mode 0.7\% ($5\times10^{32}$\,erg~s$^{-1}$) of the spin-down power \citep{BAB:2015}. Finally the optical luminosity is about 1.4\% ($10^{33}$\,erg~s$^{-1}$) of the spin-down power \citep{APS:2017}. In particular take note of the X-ray emission: matter accreting in the vicinity of the neutron star can be expected to emit most of its energy in the X-ray, so the fact that the X-ray luminosity of \psr\ is so much lower than the spin-down power suggests that physics in the vicinity of the neutron star is dominated by pulsar phenomena rather than accretion phenomena --- though the drastic change in behaviour compared to the pulsar state indicates that the matter in the accretion disk must play \emph{some} role.


During this multi-telescope campaign we observed one episode of flaring from \psr. As with previous observations, we see optical flaring simultaneous with X-ray observations. We see that every flare in all of the UV, optical and X-ray observations remains active for the same duration, indicating the possibility that each of these wavebands is related to the same underlying mechanism during flare mode events in \psr. Visual inspection of simultaneous \xmm\ and \xmm-OM observations in \citet{BPH:2014, BAB:2015, JAH:2016} confirm that every optical flare has a corresponding simultaneous X-ray flare. In contrast, radio flares have been seen to occur with no corresponding X-ray flare \citep{BDM:2018}. 

Leaving aside the time during which flares occur, we clearly demonstrate that the X-ray low and high modes have counterparts in the UV emission. This mode switching in UV happens in the same direction as that in the X-rays: in X-ray high modes the UV is brighter than in X-ray low modes. We also see mode switching in the \kep\ data (visible light) but it happens in the opposite direction: in X-ray high modes the visible light is fainter than in the X-ray low modes. 

Finally, since we know that X-ray pulsations are only present during the high modes, we folded the \hst-UV data during the high mode by using the established X-ray timing ephemeris for \psr\ and have found statistically significant millisecond pulsations at the intrinsic spin period from \psr\ in the \hst-UV data set. 

In this discussion, we focus on the broadband mode switching and pulsations from \psr, and what this teaches us about the behavior of this system during its LMXB state.  In particular, we aim to address what physical process is driving the X-ray low and high modes, and what causes the system to switch between these two regimes.

\subsection{Flaring}
In \psr, two main X-ray flare types have been observed \citep[e.g. Fig.~4 of][]{BAB:2015}: some flares are short (seconds to minutes) and isolated, and some flares appear during the prolonged flaring periods that can last for ${\gtrsim}10$\,hrs \citep{TYK:2014}. \citet{BAB:2015} see occasional intense optical and UV flares that occur in conjunction with X-ray flares of both types. In contrast, radio flaring behaviour is less directly connected to X-ray flaring, showing a variety of phenomena such as radio flares during X-ray lows, radio flares accompanying X-ray flares and isolated radio flares \citep{BDM:2018}. We also note that \citet{KCV:2018} show that the fraction of time spent in optical flaring behaviour seems to vary on a timescale of months.

Given this variety, it is not even clear whether there is a single physical mechanism or emission site that explains all the flaring behavior.  Furthermore, the observing campaign presented in this work provides only one sampling of a flaring episode, which may not be representative of flares in general.  In this one flaring episode, we do find that the substructure is similar in shape between the three bands (Fig.~\ref{fig:auto_corrs}) and effectively simultaneous (Fig.~\ref{fig:cross_corr}).

Although previous cross-correlation analysis for flare-only regions \citep{BAB:2015} have returned a variety of different lags and leads between X-ray and optical flares, \citet{BAB:2015} shows a number of flare light curves in X-ray and B-band (430~nm) where the flares appear to be simultaneous. \citet[][]{SLN:2015} seem to have observed the NIR lagging behind the optical by 300~s in non-flaring times, and \citet{HK:2018} show optical (white light) and NIR (J-band, 1200~nm) flare light curves where the flares also appear to be simultaneous. We point out that \citet{BDM:2018} observed some flares in the radio (8--12 GHz) that had no counterpart in the X-rays (and others that had). Thus it seems plausible that some but not all flares are simultaneous across a broad energy range.

The presence or absence of coherent pulsations might shed light on the origin of the flare mode, but the observational situation is not simple. \citet{ABP:2015} did not detect X-ray pulsations in the flare mode, and in fact were able to show that the absolute intensity of the pulsations must decrease during flares. \citet{PAS:2019} detect optical pulsations during the flare mode, but report that the intensity of these pulsations decreases by a factor of a few during these modes. We are unable to detect ultraviolet pulsations during the flare modes, but given the increased background and decreased signal, this is consistent with the idea that the pulsation mechanism is the same in all three bands. One possibility is that the flares are produced by a mechanism independent from the high-low mode switching, that is, when flares are occurring the high-low mode switching continues undisturbed, but some broad-band emission process also occurs. The decreased pulsed flux during flares compared to the high mode might be explained by our inability to remove low modes that occur during flaring, though as low modes only occupy ${\sim}20\%$ of the time this is not sufficient to explain the size of the pulsed flux decrease observed by \citet{PAS:2019}.

Given that we have only a single example drawn from the diverse population of flares, we cannot usefully address any explanation for the multi-wavelength flaring activity. We refer the reader to \citet{BAB:2015, BDM:2018,PRC:2018} for discussions on this topic. 


\subsection{Moding}
Unlike the erratic flaring, what we observe during the X-ray low and high mode times is more likely to be representative of an ongoing low-level accretion phenomenon in tMSPs. The moding behaviour has displayed remarkably consistent phenomenology throughout the LMXB state. The mode switches are rapid (${\sim}$few seconds) and top-hat, steady light curves persist in an extremely stable manner on multi-year timescales \citep{ABP:2015, BAB:2015, Papitto:2013, JAH:2016}. The modes do appear to directly connect to pulsations as the X-ray pulsations turn on and off with the switches between high and low modes, respectively \citep{ABP:2015}.  This suggests that the mode switches affect matter flow in the immediate vicinity of the neutron star. 

\subsubsection{Broadband spectral behaviour}
Previous studies using simultaneous, multi-wavelength observations have tried to investigate if, i) the moding behaviour extends to other frequency bands, and ii) if it is correlated with X-ray moding \citep[e.g.,][]{BAB:2015,BDM:2018,SDG:2018,Coti:2018,SLN:2015,SDG:2018,PAS:2019}. 
A significant effort has been invested in dissecting what truly is a multiwavelength problem. However, most of these studies have been limited to at most a few simultaneous bands.  Given the individual complexities of the system, a single coherent picture has been elusive. Coupled with these studies, our campaign demonstrates that synchronized moding is observable across the electromagnetic spectrum, from ${\sim} 10$\,GHz radio frequencies up to at least ${\sim} 80$\,keV hard X-ray energies.

The stability of the high/low mode dichotomy in \psr\ allows us try and join all the observational evidence into a broadband phenomenological behaviour, which we attempt to summarize below.

\begin{figure*}
\plotone{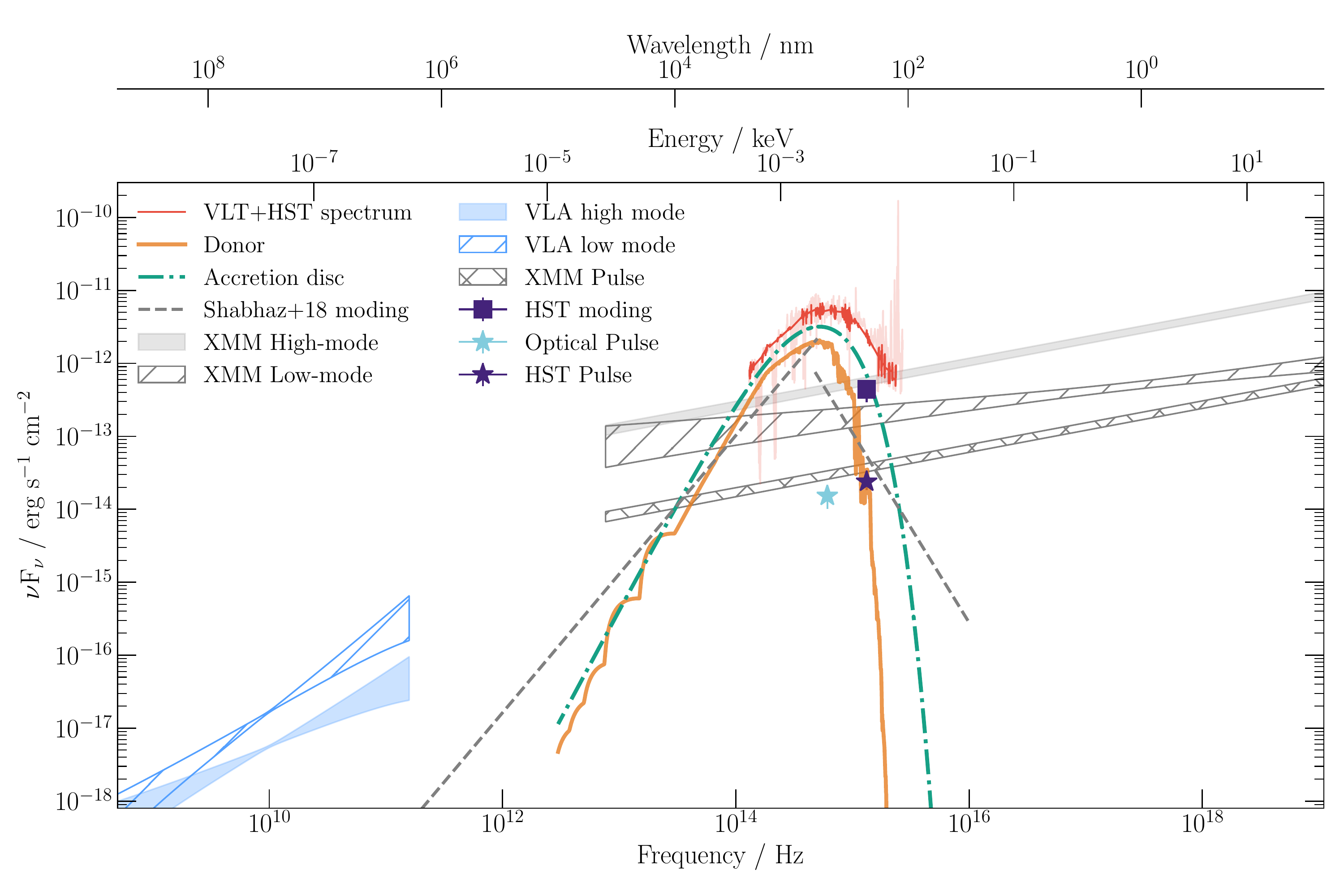}
\caption{The de-reddened broadband spectral energy distribution of \psr. Low and high modes labels are defined by the X-ray band. The extrapolated component for both X-ray modes and the pulse fractions are taken from \citet{BAB:2015} and \citet{JAH:2016}, respectively. The X-ray pulse spectrum has been assumed to be the same as for the high mode. We show the optical \citep{APS:2017} and UV pulse fractions, the latter obtained in Sec.~\ref{subsec:da}. The donor and accretion disk contribution are adapted from \citet{Her:2016}; Hern\'andez~Santisteban et al. (in prep). Radio spectrum for high- and low-modes are taken from \citet{BAB:2015}. The moding fraction observed at optical and NIR bands is adapted from \citet{SDG:2018}.
\label{fig:sed}}
\end{figure*}

\textit{The high-energy power-law} which extends from hard X-rays down to NIR, has now been confirmed by the present study, which shows UV, soft and hard X-rays to mode synchronously. Previous broadband studies had suggested that the X-ray power-law component should extend down to at least the UV band in order to explain the excess observed in the photometric data \citep{BDC:2016,Coti:2018,PAS:2019}. This was further confirmed by the SED modelling of quasi-simultaneous \hst\ and VLT/X-Shooter data \citep[][Hern\'andez Santisteban et al., in prep]{Her:2016}. We have extended this work in Fig.~\ref{fig:sed}, adding an extrapolated power-law component as measured in the X-rays \citep[using a photon index $\Gamma=1.71\pm0.01$ and $\Gamma=1.80\pm0.05$ for high- and low-mode respectively;][]{BAB:2015}. It is remarkable that the HST moding fraction (${\sim}25$\%, see Fig.~\ref{fig:mode_switches}) lies close to the expected contribution of the X-ray component in the UV. This is also supported by the turnover to higher energies observed in the far-UV data \citep[$\lesssim 140$ nm,][]{Her:2016}. Both observational facts strongly suggest that a single component indeed connects the high-energy spectrum of the mode switching down to at least UV bands. 
At the higher end, the scarcity of $\gamma$-ray photons makes it impossible to determine if the moding behaviour extends to the GeV energies (Maatman et al., \textit{in prep}). However, we know that overall emission extends to GeV ($\gamma$-rays), with an apparent crossover between hard X-ray power law and $\gamma$-ray power law at $\sim 1-10$\,MeV \citet{TYK:2014}.
At the lower energies, however \citet[][see their Fig.~5]{PAS:2019} surprisingly showed NIR $K-$band photometry with simultaneous \xmm\ where both bands present correlated behaviour. This shows then the same spectral component spans from hard X-rays to NIR. 

However, around the peak of the optical range of the SED (500-600 nm), we observe an anti-correlation between the X-rays. Previous studies already showed evidence of this fact, as seen in the bi-modal distribution of fluxes in $r'$-band (612\,nm) is inverted in reference to the X-rays \citep{SDG:2018}\footnote{In \citet{SDG:2018}, the modes are referred to as passive- and active-states, which are fainter and brighter respectively in both bands. We suggest that their passive states correspond to our (X-ray-defined) high modes, and their active states correspond to our low modes.} and in the overall analysis of the \kep\ light curve, where they show the same inverted flux distribution \citep{KCV:2018}. We argue that these inverted distributions are in fact the anti-correlated mode switches, that we observe to change simultaneously (albeit in opposite direction) with the X-rays as presented in this study (see Figure.~\ref{fig:mode_switches}).

As we go to wavelengths longer than NIR bands, the high-energy power-law has to turn over in order for the bright flat-spectrum radio emission to show anti-correlated behaviour during X-ray moding \citep{BDM:2018}. Whether this radio component and the optical component are part of the same mechanism or not, it is intriguing that they show a similar behaviour. In other low-mass X-ray binaries, jet emission can extend and peak in the optical/NIR both in high and low (quiescent) accretion rates, similar to those in \psr. This would provide a natural mechanism to explain the anti-correlated component. However, this two components could be originating from two different sites, as suggested by \citet{SDG:2018} where the optical emission arises from reprocessing emission and the NIR from synchrotron emission in the outflow .

Finally, we note that measuring moding behaviour in UV, optical, and NIR is somewhat complicated by the presence of emission from the companion and accretion disk, which are modulated at the orbital period and heavily dilutes the mode-related signal. Further observations at longer wavelengths such as far/mid-infrared and sub-mm might provide additional evidence to the origin of this component.

%

\subsection{Pulsations}

Coherent pulsations from \psr\ in the LMXB state have now been observed in X-rays \citep[$0.3-10$\,keV][]{ABP:2015}, optical \citep[$320-900$\,nm][]{APS:2017}, and UV ($180-280$\,nm). In all three cases, the pulse profile is double-peaked, that is, it is dominated by the first overtone. \citet{APS:2017} argue, based on model-derived spectral estimates, that the optical pulsations must be produced by a different physical process than the X-ray, but our results suggest a different interpretation: the pulsed fluxes in optical, UV, and X-ray bands all lie on a line (see, Fig.\ref{fig:sed}) with slope of $1.4\pm 0.6$, which is consistent with the X-ray high-mode power-law index of $1.70\pm0.05$. It is therefore plausible that the pulsations in these three bands are produced by a single process, and in fact that process also produces the mode switching. The link to mode switching is strengthened by the result of \citet{ABP:2015} showing that X-ray pulsations disappear in both flare and low modes. We do not have the statistics to verify such a disappearance for the UV pulsations, and \citet{APS:2017} do not have the mode classifications to verify this for the optical. Even stronger evidence for a broad-band emission process could be provided by showing phase alignm
ent between optical or UV pulsations and X-ray, but due to clock limitations aboard \hst\ this would require careful combination of a ground-based optical observation like that of \citet{APS:2017} with a quasi-simultaneous \xmm\ observation to constrain the pulsar ephemeris. A search for $\gamma$-ray pulsations would help explore the limits of the broad-band nature of the pulsation mechanism. However, a possible spectral break in the overall emission near $1-10$\,MeV \citep{TYK:2014} constrains the pulsations to also have a break at some similar energy.  It is thus unclear whether to expect $\gamma$-ray pulsations (Jaodand et al., \textit{in prep}) or what their pulsed fraction should be, though sampling this very different regime would be a valuable physical probe. 

\subsection{Mechanisms}

In spite of the broad range of observational data on \psr, the underlying physics remains puzzling. In addition to orbitally-modulated or un-modulated emission from the companion and outer regions of the accretion flow, it appears that there are two mechanisms at play, with rapid switching between them.  In the high mode, we get bright but stable X-ray emission and pulsations that extend from optical through UV up to X-ray. In the low mode, we get variable but bright radio emission as well as (unpulsed) emission extending from the infrared up to the optical. There is also the matter of flares, but we will leave them aside. There are three basic classes of explanation in the literature for these phenomena.

\citet{ABP:2015}, when they first noticed the mode switching, proposed that the high modes were accretion onto the surface, while the low modes were ejection of material by the propeller action of the pulsar's statically rotating magnetic field. \citet{PT:2015} also discuss this origin for the modes. Motion of the inner edge back and forth across the corotation radius then neatly explains the switching between discrete modes, at least in a simplistic picture of dynamics inside the magnetosphere. The propeller mode can be expected to eject material from the system, which essentially predicted the discovery by \citet{BDM:2018} that the low modes seemed to be accompanied by outflows. Nevertheless from the beginning it was clear that the accretion rate posed a problem: the high-mode X-ray luminosity provides an upper limit on the amount of material that can be reaching the neutron-star surface, and this amount of material is orders of magnitude too small for its ram pressure to overcome the pulsar's static magnetic pressure at or near the corotation radius. More, as we discussed above, \citet{JAH:2016} showed that the power budget in the vicinity of the neutron star was dominated by spin-down power rather than accretion power. This is in stark contrast to standard accreting millisecond X-ray pulsars (AMXPs), where channeled accretion onto the surface is indeed thought to be the origin of their (much brighter, but also unstable in phase) X-ray pulsations.

\citet{VNB:2019} describe an alternative scenario for mode production. They suggest that the low modes occur when the accreting material penetrates the light cylinder. \citet{PSB:2015} and \citet{PT:2017} argue that this should result in an increased number of open field lines and therefore an enhanced pulsar wind. This enhanced pulsar wind would then strip baryons from the accretion flow to produce a radio-bright outflow. The high modes, in this model, occur when no material enters the pulsar's light cylinder, so the pulsar's spin-down and thus wind power is unaffected. This model suggests that the enhanced X-rays in this mode come from bombardment of the inner edge of the accretion flow by the pulsar wind (which in the low modes would already be divided by the conductive material in the light cylinder). The pulsations would then be produced by the inner regions of the pulsar wind sweeping around the inner edge of the accretion flow. This mechanism is difficult in a number of ways. Very few models of non-aligned rotating pulsars launching winds exist, and the physics is quite uncertain even before including the interactions with an accretion flow near to the light cylinder. In particular it is far from established that the energy flux --- whether leptons rest mass, bulk kinetic energy, thermal, or Poynting --- should be strongly focused near the \emph{magnetic} equator, as needed to produce two hotspots on the inner face of the accretion flow. More, if the inner edge of the moves in or out, the pulses should be advanced or retarded by the light travel time. The extreme stability of the pulse phase was essential to \citet{JAH:2016} --- aside from the orbital variations, which have a very different signature, on long timescales the pulses drift by less than $5\,\mu\text{s}$, which means that the inner edge of the disk cannot move by more than about $1.5\,\text{km}$. Of course this is on average over days to years. Shorter-term variations in the inner edge would appear as smearing of the pulse profile; given the simple pulse shapes this would simply decrease the observed amplitude of each harmonic in the pulse profile. Even so, variations larger than about an eighth of a turn would pose problems, and this is only $50\,\text{km}$, less than a light cylinder radius. Since the model relies upon the inner edge to cross into the light cylinder during low modes, any such mechanism has to hold the edge of the accretion flow steady and close to the light cylinder --- but in general accreting sources are observed to have very substantial X-ray variations that presumably track substantial variations in the rate of transfer of material into the inner disk. 

\citet{ABP:2015} also pointed out a phenomenological similarity between X-ray mode switches in \psr\ and those in isolated slow radio pulsars that undergo "mode switching", a phenomenon in which radio pulse profiles change between several stable configurations \citep{B:1970}. \citet{HHK:2013} showed that these mode changes also alter the X-ray luminosity of the pulsar, and \citet{LHK:2010} showed that they also affect the spin-down rate of the pulsar. All this argues that even in isolation pulsar magnetospheres can have multiple quasi-stable configurations and can abruptly --- on second timescales --- switch between them. Such behaviour was not observed, in spite of long-term monitoring, in \psr\ while it was in the radio pulsar state, but it is possible that the observed accretion flow could somehow induce transitions in the pulsar magnetosphere. This would require material to enter the light cylinder, but the X-ray pulsations would then be generated in the magnetosphere by the pulsar mechanism. Outflows would be driven by the pulsar wind, whose intensity could plausibly change between modes (as it absorbs the majority of the spin-down power), and this wind could strip baryons from the accretion flow to produce the radio-bright outflow we see in the low modes. Unfortunately pulsar moding is very poorly understood even in isolated pulsars, let alone with the injection of accreted material. But some of the most difficult-to-explain behaviour of \psr\ is seen in isolated mode-switching pulsars: rapid (second-scale) switching between quasi-stable modes that persist for minutes to hours, associated changes in X-ray luminosity and wind, modest changes in spin-down, and stable X-ray pulsations.

The X-ray pulsations of \psr\ are only detected in the high X-ray mode \citep{ABP:2015}, and optical pulsations are detected both in the high X-ray mode and during flares \citep[where the pulsed flux is reduced by a third,][]{PAS:2019}. Presumably a similar moding behaviour also occurs for the UV pulsations we present here, though the low pulsed fraction limits us from meaningfully excluding UV pulsations during the X-ray low mode.  Various explanations have been proposed for the moding of pulsations \citep[e.g.][]{PAS:2019,VNB:2019}; here we remark that (apparently) isolated, rotation-powered pulsars are also known to display broad-band mode switching \citep{HHK:2013,MKT:2016,HKB:2018}.  We thus speculate \citep[as was also discussed in \S11.2 of][]{ABP:2015} that it may be possible that the moding behavior seen in \psr\ is an accretion-induced version of the same moding phenomenon seen in isolated pulsars.  Small amounts of hadronic material could enter into the light cylinder and cause a magnetospheric reconfiguration and enhancement of the rotation-powered pulsation mechanism.  However, given that the origin of moding in isolated pulsars remains unknown, it is hard to see how to test this scenario and, ultimately, it remains a speculative suggestion.  Nonetheless, an advantage of this model is that it can naturally explain the stability of the multi-frequency pulse profile shape and its timing stability - i.e. that the emitting region is reproducible over long timescales, which seems harder to achieve in models in which the pulsar wind, interacting with the inner edge of the accretion flow, is responsible for the pulsations.

\section{Conclusions}

We have established that the mode switching exhibited by \psr\ extends across eight decades in frequency, but that the direction of the mode switching changes somewhere in the optical band. We suggest a physical interpretation: high modes occur when accreting matter reaches into the light cylinder and possibly down to the neutron star surface (inflow), and low modes occur when instead matter is ejected in a possibly collimated outflow. This view explains why in the high mode we see the appearance of pulsations and high-energy brightness. This view also explains why in the low mode we see faint high energy emission, the disappearance of pulsations, and the emergence of flat-spectrum radio emission as well as increased NIR emission. The flare mode mechanism we leave unexplained. We note that the mode switches happen too rapidly to be directly controlled by variations in the mass transfer from the companion; turbulent variations in the disk are a more likely direct cause. Precisely what changes when the switch from inflow to outflow occurs is not obvious; we discuss three scenarios.


First, a simplistic toy model can explain the mode switching with an accretion disk terminated near the co-rotation radius by magnetospheric pressure. When the disk extends inside the corotation radius, coupling with the magnetic field can then lead to channeled accretion onto the neutron-star surface, producing a structured inflow. When the ram pressure drops, the inner edge of the disk recedes outside the corotation radius, and propeller-mode accretion takes over, stopping accretion onto the surface and replacing it with an outflow. This model is simple but the observed X-ray luminosities indicate far too little material is entering the light cylinder to produce the needed ram pressure to reach corotation in the high mode. The toy model of propeller-mode accretion is also too simplistic to work theoretically; magnetohydrodynamic simulations \citep{2004ApJ...616L.151R} actually predict mixed inflows and outflows.



Another means of switching between inflow and outflow is by having an accretion disk terminate near the light cylinder. Recently, \citet{PT:2017} put forth results from the first ever fully-relativistic MHD simulation where they model magnetized accretion onto millisecond pulsars with a force-free magnetosphere. They are able to show four distinct states depending on the accretion rate. This includes an intermediate state where the accretion disk is usually inside the light cylinder radius but at other times recedes and leads to synchrotron radiation from inflowing material being expelled in a classical `propeller' ejection. This model improves upon the simplistic model by requiring much smaller amounts of material to reach the neutron-star surface. However, here the modelling is limited to $\sim$1\,s compared to the observed mode switching time scales of a few seconds. 

Lastly, prior \hst\ observations suggest that the accretion disk may actually terminate at ${\sim}100\times$ the light cylinder radius \citep[][Hern\'andez Santisteban et al., in prep]{Her:2016}. At this distance, the light crossing time is ${\sim}60$ ms, so coherent pulsations are only possible if during inflow material reaches down near the light cylinder. This material would be opposed by the equatorial part of the pulsar wind, but such intrusions may be possible. Outflows would easily be driven by the pulsar wind, which in isolated pulsars shows well-collimated jets along the pulsar's rotation axis \citep[e.g.][]{BN:2011}. In this scenario the majority of material transferred from the companion is stripped away by the active pulsar wind, explaining why the X-ray luminosity appears very low for the amount of disk material inferred from the optical spectrum (Hern\'andez Santisteban et al., in prep). Switching between the modes could be expected to occur on the viscous timescale at the inner edge of the disk, which is a few seconds, in accordance with the observed mode switching time. The pulsar would also spend most of its time (that is, when there is no flaring) in a more or less undisturbed radio pulsar state, in accordance with the unchanged spin-down torque. The absence of detectable radio pulsations is easily explained by the presence of enough ionized material in and near the system to screen any radio pulsations from view.

In summary, we propose that the observed UV pulsations, moding and turnover in the SED of \psr\ can be explained by an `inflow-outflow' mechanism where the high modes arise from incursions of inflowing material from an accretion disk situated at ${\sim}100\times$ the light cylinder radius and the low modes arise from pulsar-wind-driven outflows.














\acknowledgments

We thank Sergio Campana, James Miller-Jones, Alessandro Papitto, Rudy Wijnands, Jakob van den Eijnden and Francesco Coti Zelati for interesting discussions related to \psr\ in its current LMXB state.  AJ and JWTH acknowledge funding from the European Research Council under the European Union's Seventh Framework Programme (FP/2007-2013) / ERC Starting Grant agreement nr. 337062 (``DRAGNET'').  JWTH also acknowledges funding from an NWO Vidi fellowship. JVHS and ND are supported by a Vidi grant from NWO, awarded to ND. SB was funded in part by NASA through grant number HST-GO-14934.002-A from the Space Telescope Science Institute (STScI), which is operated by the Association of Universities for Reasearch in Astronomy, Inc., under NASA contract NAS 5-26555. The authors are grateful to Fiona Harrison and the \nustar\ team for making the DDT observations of \psr\ possible.  This work was based in part on observations made with the NASA/ESA Hubble Space Telescope. This research has made use of data and software provided by the High Energy Astrophysics Science Archive Research Center (HEASARC), which is a service of the Astrophysics Science Division at NASA/GSFC and the High Energy Astrophysics Division of the Smithsonian Astrophysical Observatory. This work has made extensive use of NASA's Astrophysics Data System Bibliographic Services (ADS) and the arXiv.  

%

\vspace{5mm}
\facilities{\textit{HST}, \textit{Kepler}, \textit{NuSTAR}, \textit{XMM-Newton}}

\vspace{7mm}
This research made use of following community-developed software products.  
\software{\textit{Astropy} \citep{Astropy:2018}, \textit{Matplotlib} \citep{Hunter:2007}, \textit{Seaborn} \citep{Seaborn} and \textit{Scipy} \citep{Scipy}}\\

\bibliography{ApJ_ADJ}

\begin{thebibliography}{}
\expandafter\ifx\csname natexlab\endcsname\relax\def\natexlab#1{#1}\fi
\providecommand{\url}[1]{\href{#1}{#1}}
\providecommand{\dodoi}[1]{doi:~\href{http://doi.org/#1}{\nolinkurl{#1}}}
\providecommand{\doeprint}[1]{\href{http://ascl.net/#1}{\nolinkurl{http://ascl.net/#1}}}
\providecommand{\doarXiv}[1]{\href{https://arxiv.org/abs/#1}{\nolinkurl{https://arxiv.org/abs/#1}}}

\bibitem[{{Aliu} {et~al.}(2016){Aliu}, {Archambault}, {Archer}, {Benbow},
  {Bird}, {Biteau}, {Buchovecky}, {Buckley}, {Bugaev}, {Byrum}, {Cardenzana},
  {Cerruti}, {Chen}, {Ciupik}, {Connolly}, {Cui}, {Dickinson}, {Eisch},
  {Falcone}, {Feng}, {Finley}, {Fleischhack}, {Flinders}, {Fortin}, {Fortson},
  {Furniss}, {Gillanders}, {Griffin}, {Grube}, {Gyuk}, {H{\"u}tten},
  {H{\aa}kansson}, {Holder}, {Humensky}, {Johnson}, {Kaaret}, {Kar},
  {Kelley-Hoskins}, {Kertzman}, {Kieda}, {Krause}, {Lang}, {Loo}, {Maier},
  {McArthur}, {McCann}, {Meagher}, {Moriarty}, {Mukherjee}, {Nguyen}, {Nieto},
  {O'Faol{\'a}in de Bhr{\'o}ithe}, {Ong}, {Otte}, {Pandel}, {Park}, {Pelassa},
  {Petrashyk}, {Pohl}, {Popkow}, {Pueschel}, {Quinn}, {Ragan}, {Reynolds},
  {Richards}, {Roache}, {Rulten}, {Santander}, {Sembroski}, {Shahinyan},
  {Smith}, {Staszak}, {Telezhinsky}, {Tucci}, {Tyler}, {Varlotta}, {Vincent},
  {Wakely}, {Weiner}, {Weinstein}, {Wilhelm}, {Williams}, {Zitzer},
  {Chernyakova}, \& {Roberts}}]{Ver:2016}
{Aliu}, E., {Archambault}, S., {Archer}, A., {et~al.} 2016, \apj, 831, 193,
  \dodoi{10.3847/0004-637X/831/2/193}

\bibitem[{{Ambrosino} {et~al.}(2017){Ambrosino}, {Papitto}, {Stella}, {Meddi},
  {Cretaro}, {Burderi}, {Di Salvo}, {Israel}, {Ghedina}, {Di Fabrizio}, \&
  {Riverol}}]{APS:2017}
{Ambrosino}, F., {Papitto}, A., {Stella}, L., {et~al.} 2017, Nature Astronomy,
  1, 854, \dodoi{10.1038/s41550-017-0266-2}

\bibitem[{{Archibald} {et~al.}(2010){Archibald}, {Kaspi}, {Bogdanov},
  {Hessels}, {Stairs}, {Ransom}, \& {McLaughlin}}]{AKB:2010}
{Archibald}, A.~M., {Kaspi}, V.~M., {Bogdanov}, S., {et~al.} 2010, \apj, 722,
  88, \dodoi{10.1088/0004-637X/722/1/88}

\bibitem[{{Archibald} {et~al.}(2013){Archibald}, {Kaspi}, {Hessels},
  {Stappers}, {Janssen}, \& {Lyne}}]{AKH:2013}
{Archibald}, A.~M., {Kaspi}, V.~M., {Hessels}, J.~W.~T., {et~al.} 2013, ArXiv
  e-prints.
\newblock \doarXiv{1311.5161}

\bibitem[{{Archibald} {et~al.}(2015){Archibald}, {Bogdanov}, {Patruno},
  {Hessels}, {Deller}, {Bassa}, {Janssen}, {Kaspi}, {Lyne}, {Stappers},
  {Tendulkar}, {D'Angelo}, \& {Wijnands}}]{ABP:2015}
{Archibald}, A.~M., {Bogdanov}, S., {Patruno}, A., {et~al.} 2015, \apj, 807,
  62, \dodoi{10.1088/0004-637X/807/1/62}

\bibitem[{{Astropy Collaboration} {et~al.}(2018){Astropy Collaboration},
  {Price-Whelan}, {Sip{\H o}cz}, {G{\"u}nther}, {Lim}, {Crawford}, {Conseil},
  {Shupe}, {Craig}, {Dencheva}, {Ginsburg}, {VanderPlas}, {Bradley},
  {P{\'e}rez-Su{\'a}rez}, {de Val-Borro}, {Aldcroft}, {Cruz}, {Robitaille},
  {Tollerud}, {Ardelean}, {Babej}, {Bach}, {Bachetti}, {Bakanov}, {Bamford},
  {Barentsen}, {Barmby}, {Baumbach}, {Berry}, {Biscani}, {Boquien}, {Bostroem},
  {Bouma}, {Brammer}, {Bray}, {Breytenbach}, {Buddelmeijer}, {Burke},
  {Calderone}, {Cano Rodr{\'{\i}}guez}, {Cara}, {Cardoso}, {Cheedella},
  {Copin}, {Corrales}, {Crichton}, {D'Avella}, {Deil}, {Depagne}, {Dietrich},
  {Donath}, {Droettboom}, {Earl}, {Erben}, {Fabbro}, {Ferreira}, {Finethy},
  {Fox}, {Garrison}, {Gibbons}, {Goldstein}, {Gommers}, {Greco}, {Greenfield},
  {Groener}, {Grollier}, {Hagen}, {Hirst}, {Homeier}, {Horton}, {Hosseinzadeh},
  {Hu}, {Hunkeler}, {Ivezi{\'c}}, {Jain}, {Jenness}, {Kanarek}, {Kendrew},
  {Kern}, {Kerzendorf}, {Khvalko}, {King}, {Kirkby}, {Kulkarni}, {Kumar},
  {Lee}, {Lenz}, {Littlefair}, {Ma}, {Macleod}, {Mastropietro}, {McCully},
  {Montagnac}, {Morris}, {Mueller}, {Mumford}, {Muna}, {Murphy}, {Nelson},
  {Nguyen}, {Ninan}, {N{\"o}the}, {Ogaz}, {Oh}, {Parejko}, {Parley}, {Pascual},
  {Patil}, {Patil}, {Plunkett}, {Prochaska}, {Rastogi}, {Reddy Janga},
  {Sabater}, {Sakurikar}, {Seifert}, {Sherbert}, {Sherwood-Taylor}, {Shih},
  {Sick}, {Silbiger}, {Singanamalla}, {Singer}, {Sladen}, {Sooley},
  {Sornarajah}, {Streicher}, {Teuben}, {Thomas}, {Tremblay}, {Turner},
  {Terr{\'o}n}, {van Kerkwijk}, {de la Vega}, {Watkins}, {Weaver}, {Whitmore},
  {Woillez}, {Zabalza}, \& {Astropy Contributors}}]{Astropy:2018}
{Astropy Collaboration}, {Price-Whelan}, A.~M., {Sip{\H o}cz}, B.~M., {et~al.}
  2018, \aj, 156, 123, \dodoi{10.3847/1538-3881/aabc4f}

\bibitem[{{Backer}(1970)}]{B:1970}
{Backer}, D.~C. 1970, \nat, 228, 1297, \dodoi{10.1038/2281297a0}

\bibitem[{{Baglio} {et~al.}(2016){Baglio}, {D'Avanzo}, {Campana}, {Coti
  Zelati}, {Covino}, \& {Russell}}]{BDC:2016}
{Baglio}, M.~C., {D'Avanzo}, P., {Campana}, S., {et~al.} 2016, \aap, 591, A101,
  \dodoi{10.1051/0004-6361/201628383}

\bibitem[{{Bassa} {et~al.}(2014){Bassa}, {Patruno}, {Hessels}, {Keane},
  {Monard}, {Mahony}, {Bogdanov}, {Corbel}, {Edwards}, {Archibald}, {Janssen},
  {Stappers}, \& {Tendulkar}}]{BPH:2014}
{Bassa}, C.~G., {Patruno}, A., {Hessels}, J.~W.~T., {et~al.} 2014, \mnras, 441,
  1825, \dodoi{10.1093/mnras/stu708}

\bibitem[{{Bogdanov} {et~al.}(2011){Bogdanov}, {Archibald}, {Hessels}, {Kaspi},
  {Lorimer}, {McLaughlin}, {Ransom}, \& {Stairs}}]{BAH:2011}
{Bogdanov}, S., {Archibald}, A.~M., {Hessels}, J.~W.~T., {et~al.} 2011, \apj,
  742, 97, \dodoi{10.1088/0004-637X/742/2/97}

\bibitem[{{Bogdanov} {et~al.}(2005){Bogdanov}, {Grindlay}, \& {van den
  Berg}}]{BGV:2005}
{Bogdanov}, S., {Grindlay}, J.~E., \& {van den Berg}, M. 2005, \apj, 630, 1029,
  \dodoi{10.1086/432249}

\bibitem[{{Bogdanov} \& {Halpern}(2015)}]{BH:2015}
{Bogdanov}, S., \& {Halpern}, J.~P. 2015, \apjl, 803, L27,
  \dodoi{10.1088/2041-8205/803/2/L27}

\bibitem[{{Bogdanov} {et~al.}(2015){Bogdanov}, {Archibald}, {Bassa}, {Deller},
  {Halpern}, {Heald}, {Hessels}, {Janssen}, {Lyne}, {Mold{\'o}n}, {Paragi},
  {Patruno}, {Perera}, {Stappers}, {Tendulkar}, {D'Angelo}, \&
  {Wijnands}}]{BAB:2015}
{Bogdanov}, S., {Archibald}, A.~M., {Bassa}, C., {et~al.} 2015, \apj, 806, 148,
  \dodoi{10.1088/0004-637X/806/2/148}

\bibitem[{{Bogdanov} {et~al.}(2018){Bogdanov}, {Deller}, {Miller-Jones},
  {Archibald}, {Hessels}, {Jaodand}, {Patruno}, {Bassa}, \&
  {D'Angelo}}]{BDM:2018}
{Bogdanov}, S., {Deller}, A.~T., {Miller-Jones}, J.~C.~A., {et~al.} 2018, \apj,
  856, 54, \dodoi{10.3847/1538-4357/aaaeb9}

\bibitem[{{Bucciantini}(2011)}]{BN:2011}
{Bucciantini}, N. 2011, Astrophysics and Space Science Proceedings, 21, 473,
  \dodoi{10.1007/978-3-642-17251-9_39}

\bibitem[{{Coti Zelati} {et~al.}(2018){Coti Zelati}, {Campana}, {Braito},
  {Baglio}, {D'Avanzo}, {Rea}, \& {Torres}}]{Coti:2018}
{Coti Zelati}, F., {Campana}, S., {Braito}, V., {et~al.} 2018, \aap, 611, A14,
  \dodoi{10.1051/0004-6361/201732244}

\bibitem[{{de Martino} {et~al.}(2010){de Martino}, {Falanga}, {Bonnet-Bidaud},
  {Belloni}, {Mouchet}, {Masetti}, {Andruchow}, {Cellone}, {Mukai}, \&
  {Matt}}]{deMartino:2010}
{de Martino}, D., {Falanga}, M., {Bonnet-Bidaud}, J.-M., {et~al.} 2010, \aap,
  515, A25, \dodoi{10.1051/0004-6361/200913802}

\bibitem[{{de Martino} {et~al.}(2013){de Martino}, {Belloni}, {Falanga},
  {Papitto}, {Motta}, {Pellizzoni}, {Evangelista}, {Piano}, {Masetti},
  {Bonnet-Bidaud}, {Mouchet}, {Mukai}, \& {Possenti}}]{deMartino:2013}
{de Martino}, D., {Belloni}, T., {Falanga}, M., {et~al.} 2013, \aap, 550, A89,
  \dodoi{10.1051/0004-6361/201220393}

\bibitem[{{Deller} {et~al.}(2012){Deller}, {Archibald}, {Brisken},
  {Chatterjee}, {Janssen}, {Kaspi}, {Lorimer}, {Lyne}, {McLaughlin}, {Ransom},
  {Stairs}, \& {Stappers}}]{DAB:2012}
{Deller}, A.~T., {Archibald}, A.~M., {Brisken}, W.~F., {et~al.} 2012, \apjl,
  756, L25, \dodoi{10.1088/2041-8205/756/2/L25}

\bibitem[{{Deller} {et~al.}(2015){Deller}, {Moldon}, {Miller-Jones}, {Patruno},
  {Hessels}, {Archibald}, {Paragi}, {Heald}, \& {Vilchez}}]{DMM:2015}
{Deller}, A.~T., {Moldon}, J., {Miller-Jones}, J.~C.~A., {et~al.} 2015, \apj,
  809, 13, \dodoi{10.1088/0004-637X/809/1/13}

\bibitem[{{Eracleous} \& {Horne}(1996)}]{Eracleous:1996aa}
{Eracleous}, M., \& {Horne}, K. 1996, \apj, 471, 427

\bibitem[{{Frank} {et~al.}(2002){Frank}, {King}, \& {Raine}}]{FAR:2002}
{Frank}, J., {King}, A., \& {Raine}, D.~J. 2002, {Accretion Power in
  Astrophysics: Third Edition}, 398

\bibitem[{Hakala \& Kajava(2018)}]{HK:2018}
Hakala, P., \& Kajava, J. J.~E. 2018, Monthly Notices of the Royal Astronomical
  Society, 474.
\newblock \url{http://dx.doi.org/10.1093/mnras/stx2922}

\bibitem[{{Halpern} {et~al.}(2013){Halpern}, {Gaidos}, {Sheffield},
  {Price-Whelan}, \& {Bogdanov}}]{HGS:2013}
{Halpern}, J.~P., {Gaidos}, E., {Sheffield}, A., {Price-Whelan}, A.~M., \&
  {Bogdanov}, S. 2013, The Astronomer's Telegram, 5514

\bibitem[{{Harrison} {et~al.}(2013){Harrison}, {Craig}, {Christensen},
  {Hailey}, {Zhang}, {Boggs}, {Stern}, {Cook}, {Forster}, {Giommi},
  {Grefenstette}, {Kim}, {Kitaguchi}, {Koglin}, {Madsen}, {Mao}, {Miyasaka},
  {Mori}, {Perri}, {Pivovaroff}, {Puccetti}, {Rana}, {Westergaard}, {Willis},
  {Zoglauer}, {An}, {Bachetti}, {Barri{\`e}re}, {Bellm}, {Bhalerao},
  {Brejnholt}, {Fuerst}, {Liebe}, {Markwardt}, {Nynka}, {Vogel}, {Walton},
  {Wik}, {Alexander}, {Cominsky}, {Hornschemeier}, {Hornstrup}, {Kaspi},
  {Madejski}, {Matt}, {Molendi}, {Smith}, {Tomsick}, {Ajello}, {Ballantyne},
  {Balokovi{\'c}}, {Barret}, {Bauer}, {Blandford}, {Brandt}, {Brenneman},
  {Chiang}, {Chakrabarty}, {Chenevez}, {Comastri}, {Dufour}, {Elvis}, {Fabian},
  {Farrah}, {Fryer}, {Gotthelf}, {Grindlay}, {Helfand}, {Krivonos}, {Meier},
  {Miller}, {Natalucci}, {Ogle}, {Ofek}, {Ptak}, {Reynolds}, {Rigby},
  {Tagliaferri}, {Thorsett}, {Treister}, \& {Urry}}]{2013ApJ...770..103H}
{Harrison}, F.~A., {Craig}, W.~W., {Christensen}, F.~E., {et~al.} 2013, \apj,
  770, 103, \dodoi{10.1088/0004-637X/770/2/103}

\bibitem[{{Hermsen} {et~al.}(2013){Hermsen}, {Hessels}, {Kuiper}, {van
  Leeuwen}, {Mitra}, {de Plaa}, {Rankin}, {Stappers}, {Wright}, {Basu},
  {Alexov}, {Coenen}, {Grie{\ss}meier}, {Hassall}, {Karastergiou}, {Keane},
  {Kondratiev}, {Kramer}, {Kuniyoshi}, {Noutsos}, {Serylak}, {Pilia}, {Sobey},
  {Weltevrede}, {Zagkouris}, {Asgekar}, {Avruch}, {Batejat}, {Bell}, {Bell},
  {Bentum}, {Bernardi}, {Best}, {B{\^i}rzan}, {Bonafede}, {Breitling},
  {Broderick}, {Br{\"u}ggen}, {Butcher}, {Ciardi}, {Duscha}, {Eisl{\"o}ffel},
  {Falcke}, {Fender}, {Ferrari}, {Frieswijk}, {Garrett}, {de Gasperin}, {de
  Geus}, {Gunst}, {Heald}, {Hoeft}, {Horneffer}, {Iacobelli}, {Kuper}, {Maat},
  {Macario}, {Markoff}, {McKean}, {Mevius}, {Miller-Jones}, {Morganti}, {Munk},
  {Orr{\'u}}, {Paas}, {Pandey-Pommier}, {Pandey}, {Pizzo}, {Polatidis},
  {Rawlings}, {Reich}, {R{\"o}ttgering}, {Scaife}, {Schoenmakers}, {Shulevski},
  {Sluman}, {Steinmetz}, {Tagger}, {Tang}, {Tasse}, {ter Veen}, {Vermeulen},
  {van de Brink}, {van Weeren}, {Wijers}, {Wise}, {Wucknitz}, {Yatawatta}, \&
  {Zarka}}]{HHK:2013}
{Hermsen}, W., {Hessels}, J.~W.~T., {Kuiper}, L., {et~al.} 2013, Science, 339,
  436, \dodoi{10.1126/science.1230960}

\bibitem[{{Hermsen} {et~al.}(2018){Hermsen}, {Kuiper}, {Basu}, {Hessels},
  {Mitra}, {Rankin}, {Stappers}, {Wright}, {Grie{\ss}meier}, \&
  {Serylak}}]{HKB:2018}
{Hermsen}, W., {Kuiper}, L., {Basu}, R., {et~al.} 2018, \mnras, 480, 3655,
  \dodoi{10.1093/mnras/sty2075}

\bibitem[{{Hern\'andez Santisteban}(2016)}]{Her:2016}
{Hern\'andez Santisteban}, J.~V. 2016, PhD thesis, UNIVERSITY OF SOUTHAMPTON

\bibitem[{{Hui} \& {Becker}(2006)}]{HuiB:2006}
{Hui}, C.~Y., \& {Becker}, W. 2006, \aap, 448, L13,
  \dodoi{10.1051/0004-6361:200600008}

\bibitem[{Hunter(2007)}]{Hunter:2007}
Hunter, J.~D. 2007, Computing In Science \& Engineering, 9, 90,
  \dodoi{10.1109/MCSE.2007.55}

\bibitem[{{Jaodand} {et~al.}(2016){Jaodand}, {Archibald}, {Hessels},
  {Bogdanov}, {D'Angelo}, {Patruno}, {Bassa}, \& {Deller}}]{JAH:2016}
{Jaodand}, A., {Archibald}, A.~M., {Hessels}, J.~W.~T., {et~al.} 2016, \apj,
  830, 122, \dodoi{10.3847/0004-637X/830/2/122}

\bibitem[{{Johnson} {et~al.}(2015){Johnson}, {Ray}, {Roy}, {Cheung}, {Harding},
  {Pletsch}, {Fort}, {Camilo}, {Deneva}, {Bhattacharyya}, {Stappers}, \&
  {Kerr}}]{JRR:2015}
{Johnson}, T.~J., {Ray}, P.~S., {Roy}, J., {et~al.} 2015, \apj, 806, 91,
  \dodoi{10.1088/0004-637X/806/1/91}

\bibitem[{Jones {et~al.}(2001{\natexlab{a}})Jones, Oliphant, Peterson,
  {et~al.}}]{Seaborn}
Jones, E., Oliphant, T., Peterson, P., {et~al.} 2001{\natexlab{a}}, {SciPy}:
  Open source scientific tools for {Python}.
\newblock \url{http://www.scipy.org/}

\bibitem[{Jones {et~al.}(2001{\natexlab{b}})Jones, Oliphant, Peterson,
  {et~al.}}]{Scipy}
---. 2001{\natexlab{b}}, {SciPy}: Open source scientific tools for {Python}.
\newblock \url{http://www.scipy.org/}

\bibitem[{{Kennedy} {et~al.}(2018){Kennedy}, {Clark}, {Voisin}, \&
  {Breton}}]{KCV:2018}
{Kennedy}, M.~R., {Clark}, C.~J., {Voisin}, G., \& {Breton}, R.~P. 2018,
  \mnras, 477, 1120, \dodoi{10.1093/mnras/sty731}

\bibitem[{{Lyne} {et~al.}(2010){Lyne}, {Hobbs}, {Kramer}, {Stairs}, \&
  {Stappers}}]{LHK:2010}
{Lyne}, A., {Hobbs}, G., {Kramer}, M., {Stairs}, I., \& {Stappers}, B. 2010,
  Science, 329, 408, \dodoi{10.1126/science.1186683}

\bibitem[{{Mason} {et~al.}(2001){Mason}, {Breeveld}, {Much}, {Carter},
  {Cordova}, {Cropper}, {Fordham}, {Huckle}, {Ho}, {Kawakami}, {Kennea},
  {Kennedy}, {Mittaz}, {Pandel}, {Priedhorsky}, {Sasseen}, {Shirey}, {Smith},
  \& {Vreux}}]{MBM:2001}
{Mason}, K.~O., {Breeveld}, A., {Much}, R., {et~al.} 2001, \aap, 365, L36,
  \dodoi{10.1051/0004-6361:20000044}

\bibitem[{{McConnell} {et~al.}(2015){McConnell}, {Callanan}, {Kennedy},
  {Hurley}, {Garnavich}, \& {Menzies}}]{McConnell:2015}
{McConnell}, O., {Callanan}, P.~J., {Kennedy}, M., {et~al.} 2015, \mnras, 451,
  3468

\bibitem[{{Mereghetti} {et~al.}(2016){Mereghetti}, {Kuiper}, {Tiengo},
  {Hessels}, {Hermsen}, {Stovall}, {Possenti}, {Rankin}, {Esposito}, \&
  {Turolla}}]{MKT:2016}
{Mereghetti}, S., {Kuiper}, L., {Tiengo}, A., {et~al.} 2016, \apj, 831, 21,
  \dodoi{10.3847/0004-637X/831/1/21}

\bibitem[{{Papitto} \& {Torres}(2015)}]{PT:2015}
{Papitto}, A., \& {Torres}, D.~F. 2015, \apj, 807, 33,
  \dodoi{10.1088/0004-637X/807/1/33}

\bibitem[{{Papitto} {et~al.}(2013){Papitto}, {Hessels}, {Burgay}, {Ransom},
  {Rea}, {Possenti}, {Stairs}, {Ferrigno}, \& {Bozz}}]{Papitto:2013}
{Papitto}, A., {Hessels}, J.~W.~T., {Burgay}, M., {et~al.} 2013, The
  Astronomer's Telegram, 5069, 1

\bibitem[{{Papitto} {et~al.}(2018){Papitto}, {Rea}, {Coti Zelati}, {de
  Martino}, {Scaringi}, {Campana}, {de O{\'n}a Wilhelmi}, {Knigge},
  {Serenelli}, {Stella}, {Torres}, {D'Avanzo}, \& {Israel}}]{PRC:2018}
{Papitto}, A., {Rea}, N., {Coti Zelati}, F., {et~al.} 2018, \apjl, 858, L12,
  \dodoi{10.3847/2041-8213/aabee9}

\bibitem[{{Papitto} {et~al.}(2019){Papitto}, {Ambrosino}, {Stella}, {Torres},
  {Coti Zelati}, {Ghedina}, {Meddi}, {Sanna}, {Casella}, {Dallilar},
  {Eikenberry}, {Israel}, {Onori}, {Piranomonte}, {Bozzo}, {Burderi},
  {Campana}, {de Martino}, {Di Salvo}, {Ferrigno}, {Rea}, {Riggio}, {Serrano},
  {Veledina}, \& {Zampieri}}]{PAS:2019}
{Papitto}, A., {Ambrosino}, F., {Stella}, L., {et~al.} 2019, arXiv e-prints,
  arXiv:1904.10433.
\newblock \doarXiv{1904.10433}

\bibitem[{{Parfrey} {et~al.}(2015){Parfrey}, {Spitkovsky}, \&
  {Beloborodov}}]{PSB:2015}
{Parfrey}, K., {Spitkovsky}, A., \& {Beloborodov}, A.~M. 2015, ArXiv e-prints.
\newblock \doarXiv{1507.08627}

\bibitem[{{Parfrey} \& {Tchekhovskoy}(2017)}]{PT:2017}
{Parfrey}, K., \& {Tchekhovskoy}, A. 2017, \apj, 851, L34,
  \dodoi{10.3847/2041-8213/aa9c85}

\bibitem[{{Patruno} {et~al.}(2014){Patruno}, {Archibald}, {Hessels},
  {Bogdanov}, {Stappers}, {Bassa}, {Janssen}, {Kaspi}, {Tendulkar}, \&
  {Lyne}}]{PAH:2014}
{Patruno}, A., {Archibald}, A.~M., {Hessels}, J.~W.~T., {et~al.} 2014, \apjl,
  781, L3, \dodoi{10.1088/2041-8205/781/1/L3}

\bibitem[{{Roberts}(2013)}]{Mal:2013}
{Roberts}, M.~S.~E. 2013, in IAU Symposium, Vol. 291, IAU Symposium, ed.
  J.~{van Leeuwen}, 127--132, \dodoi{10.1017/S174392131202337X}

\bibitem[{{Romanova} {et~al.}(2004){Romanova}, {Ustyugova}, {Koldoba}, \&
  {Lovelace}}]{2004ApJ...616L.151R}
{Romanova}, M.~M., {Ustyugova}, G.~V., {Koldoba}, A.~V., \& {Lovelace},
  R.~V.~E. 2004, \apjl, 616, L151, \dodoi{10.1086/426586}

\bibitem[{{Shahbaz} {et~al.}(2018){Shahbaz}, {Dallilar}, {Garner},
  {Eikenberry}, {Veledina}, \& {Gandhi}}]{SDG:2018}
{Shahbaz}, T., {Dallilar}, Y., {Garner}, A., {et~al.} 2018, \mnras, 477, 566,
  \dodoi{10.1093/mnras/sty562}

\bibitem[{{Shahbaz} {et~al.}(2019){Shahbaz}, {Linares}, {Rodr{\'\i}guez-Gil},
  \& {Casares}}]{SLR:2019}
{Shahbaz}, T., {Linares}, M., {Rodr{\'\i}guez-Gil}, P., \& {Casares}, J. 2019,
  \mnras, 1561, \dodoi{10.1093/mnras/stz1652}

\bibitem[{{Shahbaz} {et~al.}(2015){Shahbaz}, {Linares}, {Nevado},
  {Rodr{\'{\i}}guez-Gil}, {Casares}, {Dhillon}, {Marsh}, {Littlefair},
  {Leckngam}, \& {Poshyachinda}}]{SLN:2015}
{Shahbaz}, T., {Linares}, M., {Nevado}, S.~P., {et~al.} 2015, \mnras, 453,
  3461, \dodoi{10.1093/mnras/stv1686}

\bibitem[{{Stappers} {et~al.}(2013){Stappers}, {Archibald}, {Bassa}, {Hessels},
  {Janssen}, {Kaspi}, {Lyne}, {Patruno}, \& {Hill}}]{SAB:2013}
{Stappers}, B.~W., {Archibald}, A., {Bassa}, C., {et~al.} 2013, The
  Astronomer's Telegram, 5513, 1

\bibitem[{{Stappers} {et~al.}(2014){Stappers}, {Archibald}, {Hessels}, {Bassa},
  {Bogdanov}, {Janssen}, {Kaspi}, {Lyne}, {Patruno}, {Tendulkar}, \&
  {Hill}}]{SAH:2014}
{Stappers}, B.~W., {Archibald}, A.~M., {Hessels}, J.~W.~T., {et~al.} 2014,
  \apj, 790, 39, \dodoi{10.1088/0004-637X/790/1/39}

\bibitem[{{Str{\"u}der} {et~al.}(2001){Str{\"u}der}, {Briel}, {Dennerl},
  {Hartmann}, {Kendziorra}, {Meidinger}, {Pfeffermann}, {Reppin}, {Aschenbach},
  {Bornemann}, {Br{\"a}uninger}, {Burkert}, {Elender}, {Freyberg}, {Haberl},
  {Hartner}, {Heuschmann}, {Hippmann}, {Kastelic}, {Kemmer}, {Kettenring},
  {Kink}, {Krause}, {M{\"u}ller}, {Oppitz}, {Pietsch}, {Popp}, {Predehl},
  {Read}, {Stephan}, {St{\"o}tter}, {Tr{\"u}mper}, {Holl}, {Kemmer}, {Soltau},
  {St{\"o}tter}, {Weber}, {Weichert}, {von Zanthier}, {Carathanassis}, {Lutz},
  {Richter}, {Solc}, {B{\"o}ttcher}, {Kuster}, {Staubert}, {Abbey}, {Holland},
  {Turner}, {Balasini}, {Bignami}, {La Palombara}, {Villa}, {Buttler},
  {Gianini}, {Lain{\'e}}, {Lumb}, \& {Dhez}}]{SBD:2001}
{Str{\"u}der}, L., {Briel}, U., {Dennerl}, K., {et~al.} 2001, \aap, 365, L18,
  \dodoi{10.1051/0004-6361:20000066}

\bibitem[{{Takata} {et~al.}(2014){Takata}, {Li}, {Leung}, {Kong}, {Tam}, {Hui},
  {Wu}, {Xing}, {Cao}, {Tang}, {Wang}, \& {Cheng}}]{TLK:2014}
{Takata}, J., {Li}, K.~L., {Leung}, G.~C.~K., {et~al.} 2014, \apj, 785, 131,
  \dodoi{10.1088/0004-637X/785/2/131}

\bibitem[{{Tendulkar} {et~al.}(2014){Tendulkar}, {Yang}, {An}, {Kaspi},
  {Archibald}, {Bassa}, {Bellm}, {Bogdanov}, {Harrison}, {Hessels}, {Janssen},
  {Lyne}, {Patruno}, {Stappers}, {Stern}, {Tomsick}, {Boggs}, {Chakrabarty},
  {Christensen}, {Craig}, {Hailey}, \& {Zhang}}]{TYK:2014}
{Tendulkar}, S.~P., {Yang}, C., {An}, H., {et~al.} 2014, in AAS/High Energy
  Astrophysics Division, Vol.~14, AAS/High Energy Astrophysics Division,
  122--23

\bibitem[{{Thorstensen} \& {Armstrong}(2005)}]{TA:2005}
{Thorstensen}, J.~R., \& {Armstrong}, E. 2005, \aj, 130, 759,
  \dodoi{10.1086/431326}

\bibitem[{{Turner} {et~al.}(2001){Turner}, {Abbey}, {Arnaud}, {Balasini},
  {Barbera}, {Belsole}, {Bennie}, {Bernard}, {Bignami}, {Boer}, {Briel},
  {Butler}, {Cara}, {Chabaud}, {Cole}, {Collura}, {Conte}, {Cros}, {Denby},
  {Dhez}, {Di Coco}, {Dowson}, {Ferrando}, {Ghizzardi}, {Gianotti}, {Goodall},
  {Gretton}, {Griffiths}, {Hainaut}, {Hochedez}, {Holland}, {Jourdain},
  {Kendziorra}, {Lagostina}, {Laine}, {La Palombara}, {Lortholary}, {Lumb},
  {Marty}, {Molendi}, {Pigot}, {Poindron}, {Pounds}, {Reeves}, {Reppin},
  {Rothenflug}, {Salvetat}, {Sauvageot}, {Schmitt}, {Sembay}, {Short},
  {Spragg}, {Stephen}, {Str{\"u}der}, {Tiengo}, {Trifoglio}, {Tr{\"u}mper},
  {Vercellone}, {Vigroux}, {Villa}, {Ward}, {Whitehead}, \& {Zonca}}]{TAB:2001}
{Turner}, M.~J.~L., {Abbey}, A., {Arnaud}, M., {et~al.} 2001, \aap, 365, L27,
  \dodoi{10.1051/0004-6361:20000087}

\bibitem[{{Veledina} {et~al.}(2019){Veledina}, {N{\"a}ttil{\"a}}, \&
  {Beloborodov}}]{VNB:2019}
{Veledina}, A., {N{\"a}ttil{\"a}}, J., \& {Beloborodov}, A.~M. 2019, arXiv
  e-prints, arXiv:1906.02519.
\newblock \doarXiv{1906.02519}

\bibitem[{{Wang} {et~al.}(2009){Wang}, {Archibald}, {Thorstensen}, {Kaspi},
  {Lorimer}, {Stairs}, \& {Ransom}}]{WAT:2009}
{Wang}, Z., {Archibald}, A.~M., {Thorstensen}, J.~R., {et~al.} 2009, \apj, 703,
  2017, \dodoi{10.1088/0004-637X/703/2/2017}

\bibitem[{{Wijnands} {et~al.}(2017){Wijnands}, {Degenaar}, \&
  {Page}}]{WDP:2017}
{Wijnands}, R., {Degenaar}, N., \& {Page}, D. 2017, Journal of Astrophysics and
  Astronomy, 38, 49, \dodoi{10.1007/s12036-017-9466-5}

\end{thebibliography}
\bibliographystyle{aasjournal_n}






\end{document}